\documentclass[12pt]{article}
\usepackage{xcolor,pstool,enumitem,tikz,tensor,amsthm,mathtools,soul,braket,amsbsy,amsmath,amsfonts,amssymb,cite,csquotes,mleftright}
\usepackage{bbold}  
\usepackage{verbatim} 
\DeclareMathAlphabet{\mathbbold}{U}{bbold}{m}{n}  
\DeclareMathAlphabet{\mathbb}{U}{msb}{m}{n}  
\mleftright 

\usepackage[unicode,colorlinks = true,
            linkcolor = blue!70!black,
            urlcolor  = red!70!black,
            citecolor = green!55!black,
            anchorcolor = blue!70!black]{hyperref}
\usepackage[margin=0.8in]{geometry}

\begingroup
\global\mathcode`/=\numexpr "4000 + \count2\relax
\endgroup

\setlist[enumerate]{listparindent=\parindent,parsep=0pt}

\graphicspath{{./figures/}}

\renewcommand*{\Re}{\operatorname{Re}}
\renewcommand*{\Im}{\operatorname{Im}}
\newcommand{\di}{\mathrm{d}}
\newcommand{\abs}[1]{\mathopen{|}#1\mathclose{|}}

\newcommand{\Circled}[1]{%
  \tikz[baseline=(char.base)]{%
    \node[draw,circle,inner sep=.5pt] (char) {#1};%
  }%
}

\newcommand*{\textdash}{} 
\DeclareRobustCommand{\textdash}{\unskip\nobreak\space\textendash\nobreak\space\ignorespaces}

\begin{document}
\begin{titlepage}
\vspace*{20mm}
\begin{center}
{\LARGE \bf An ode to instantons}
\vspace*{15mm}

{\large Oliver Janssen$^{1,2,3}$, Joel Karlsson$^{2,3,4}$, Flavio Riccardi$^{5,6}$ and Mattia Varrone$^{1}$}

\vspace*{8mm}

{\small
$^{1}$Laboratory for Theoretical Fundamental Physics, EPFL, 1015 Lausanne, Switzerland\\
\vspace*{2mm}
$^{2}$Institute for Theoretical Physics, KU Leuven, 3001 Leuven, Belgium\\
\vspace*{2mm}
$^{3}$Leuven Gravity Institute, KU Leuven, 3001 Leuven, Belgium\\
\vspace*{2mm}
$^{4}$DAMTP, University of Cambridge, Cambridge CB3 0WA, UK\\
\vspace*{2mm}
$^{5}$Dipartimento di Fisica, Sapienza Università di Roma, Piazzale Aldo Moro 5, 00185, Roma, Italy\\
\vspace*{2mm}
$^{6}$INFN sezione di Roma, Piazzale Aldo Moro 5, 00185, Roma, Italy
}

\vspace*{0.7cm}
\end{center}
\begin{abstract}

\noindent \normalsize We review a formalism for semiclassical time evolution in quantum mechanics, identifying complex saddle points in real time, real saddle points in complex time, and complex saddle points in complex time that reproduce known answers to classic problems. For the decay of a metastable state, we find finite-time and finite-energy analogs of the ``bounce'' which have no zero or negative modes. The one-loop phase of the wave function and the multiplicity of bounce solutions at late times are discussed. The motivation for this work is to learn how to compute decay rates in quantum field theory in situations with interesting time dependence, by first taking a humble step back to the fascinating world of quantum mechanics.

\end{abstract}

\vspace{1cm}

\begin{center}
\emph{In memory of Sidney Coleman \\ on the 89th anniversary of his birth and the 19th anniversary of his passing}
\end{center}

\vspace{0.7cm}

v1: March 7, 2026

v2: November 18, 2026

\end{titlepage}

\tableofcontents

\section{Motivation}
This work concerns the semiclassical limit in the path integral approach to quantum theory \cite{RevModPhys.20.367} \textdash a venerable topic that hardly requires an introduction \cite{Berry:1972na}. Despite its widespread use, however, some aspects of the path integral approach remain insufficiently understood.\footnote{At least, by the authors!} Especially in the context of tunneling phenomena in quantum field theory, where the path integral is our only analytic tool, and more specifically in situations with non-trivial time dependence, there are concrete, longstanding open problems, including
\begin{enumerate}
    \item What is the time-dependent decay probability $\Gamma(t)$ per unit spacetime volume of a scalar field $\phi$ in flat space, with an initial spatially homogeneous but oscillating classical field profile $\langle \phi(\boldsymbol{x},t) \rangle = \phi_\textsf{FV} + \varepsilon \cos(m t)$? Here $\phi$ is assumed subject to a potential $V$ with a metastable minimum at $\phi_\textsf{FV}$, $m^2 = V''(\phi_\textsf{FV})$ and $\varepsilon$ is a small amplitude that keeps the classical field well inside the metastable region.

    At the very least one can say that no consensus has been reached on the solution to this problem.\footnote{At least, before \cite{Janssen:2026bsi}, which appeared in between v1 and v2 of this paper.} In \cite{Keski-Vakkuri:1996lbi}, one obtains an additive correction to the undisputed $\varepsilon = 0$ result for $\log \Gamma$ of \cite{Coleman:1977py}, involving a notable exponentially enhanced factor $(\varepsilon R)^2 (m R)^{-1/2} e^{2 m R} \cos(2 m t)$, where $R$ is the nucleated bubble's radius and $1/m$ is its thickness. The calculation was done in the thin-wall limit where $m R \gg 1$. On the other hand, in \cite{Darme:2019ubo} one obtains a multiplicative correction to $\log \Gamma$ of the form $1 + (\varepsilon/\Delta \phi)^2 (m R) \cos (2m t)$, where $\Delta \phi = \abs{\phi_\textsf{TV} - \phi_\textsf{FV}}$ and $\phi_\textsf{TV}$ is the stable minimum of $V$. The derivations of \cite{Keski-Vakkuri:1996lbi,Darme:2019ubo} disagree in how the analytic continuation to imaginary time should be done, leading to discrepant results. Perhaps the most puzzling feature of both answers is that the period of the decay rate is \emph{half} that of the classical scalar's. In any case, an independent calculation would be clarifying, confirming one of the two results \textdash or none.

    \item Consider a two-field inflation model where a scalar field $\phi = (\phi_1,\phi_2)$ subject to a potential $V$ slowly rolls down the $1$-direction, with a large mass in the $2$-direction. Assume that further on in the $2$-direction, there is a region separated from the $1$-direction by a finite barrier. What is $\Gamma$ as a function of the expected value $\langle \phi_1 (t) \rangle$ of the scalar field during inflation?

    This question, and the previous one, was raised by the community around the time of \cite{PhysRevD.44.2306}. Although progress was made on when a \enquote{quasistatic} approximation is valid,\footnote{Namely, when $\partial_t \log \, \langle \phi_1 (t) \rangle \ll 1/R$. Quasistatic means we can take a snapshot of the state at time $t$ and extract $\Gamma(t)$ as if it were a static system.} a detailed answer remained lacking. More recently, \cite{Kleban:2023ugf} considered the case where the $1$- and $2$-directions are connected by a unique, regular Euclidean instanton. Denote the endpoint of the 1-side of the instanton by $\phi_{1,*}$. In this case the authors plausibly proposed that $\Gamma$ should be proportional to the product of the time-dependent value of the square of the inflaton wave functional for configurations $\phi(\boldsymbol{x})$ containing a homogeneous region of size $R$ filled with $\phi_{1,*}$, and the static result for $\Gamma$ of \cite{Coleman:1980aw} for the instanton. This leads to a short but non-vanishing time interval during which the tunneling effectively happens. Further, \cite{Kleban:2023ugf} made a second proposal in which $\Gamma$ should be calculated quasistatically by imagining a small deformation of the potential at $\langle \phi_1(t) \rangle$ that creates a local minimum there, generating a new, instantaneous instanton connecting the $1$- and $2$-directions whose action serves as an estimate for $-\log \Gamma$. They showed that the two approximation schemes for $\Gamma$ agree at tree level, up to $\mathcal{O}(1)$ factors which clearly depend on how exactly the inflaton wave functional is approximated and how the potential deformation is done. While it is hard to imagine that these proposals are qualitatively incorrect, one would like to place them on firmer footing with a first-principles calculation of the precise time dependence of $\Gamma$. Most recently, progress on this problem was made in~\cite{Yamada:2026xsd,Janssen:2026bsi}.
\end{enumerate}
Both of these problems have concrete manifestations in early universe cosmology. For problem 1, after inflation, in landscapes such as \cite{Bachlechner:2018gew} the inflaton could evolve toward a minimum with a small vacuum energy density, around which it oscillates and acts like dark matter. This minimum is metastable, however, and a decay to other regions of the landscape is possible. The oscillations enhance the decay probability by an exponential factor, which must be appropriately bounded by the age of the universe if this model is to be viable. Problem 2 has been proposed as a mechanism that could produce primordial black hole dark matter \cite{Garriga:2015fdk,Deng:2016vzb,PhysRevLett.125.181304,Kleban:2023ugf}. One dreams to go beyond plausible estimates, however, and calculate the detailed spectrum of masses of these black holes, in particular the dependence on the shape of the barrier $V(\phi_1,\phi_2)$.

To study quantum tunneling from the path integral, the standard approach of our Teacher \cite{Coleman:1978ae} is to rotate to imaginary time, and (implicitly) assume the initial state is the would-be ground state of the false vacuum. To answer questions 1 and 2 above, however, we must develop an understanding in real time and include excited states. Many groups across space and time have been interested in this problem; a biased selection of insightful work closely related to this paper include \cite{Schiff:2008heb,Andreassen:2016cvx,Mou:2019gyl,ai2019functional,Blum:2023wnb,Batini:2023zpi,Garbrecht:2024end,Garbrecht:2025tos,Lin:2025bjn,Chao:2025rhr} and references therein. Our approach will be to study analogs of questions 1 and 2 above in quantum mechanics, and, mostly, in one dimension. We will adopt a \enquote{follow-your-nose} approach, starting from the path integral expression for the time evolution of a given semiclassical initial state (\S\ref{mainsec}). This idea was introduced at least as early as \cite{Klauder1978}, and more recently continued in \cite{Schiff:2008heb,Turok:2013dfa,Blum:2023wnb}. We calculate the one-loop prefactor (agreeing with \cite{Schiff:2008heb}), up to a normalization, which we leave as an open question and speculate upon in the conclusion. We also connect with more recent discussions about which saddle points contribute to the path integral \cite{Witten:2010cx}, setting up the precise question that would determine this for our problem, but unfortunately we are not able to solve it in general. In \S\ref{examplessec} we illustrate the formalism in a set of classic problems in one-dimensional quantum mechanics, including the harmonic oscillator (\S\ref{HOsec}), the transmission of a particle with insufficient energy to surpass a barrier (\S\ref{underbarriersec}), the reflection of a particle with enough energy to surpass a barrier (\S\ref{overbarriersec}), the decay of a metastable state (\S\ref{decaysec}) and the enhanced transmission of certain particles that tunnel through multiple barriers (\S\ref{resonantsec}).

Let the reader harbor no illusion that we lay claim to revolutionary results. Indeed, similar perspectives have appeared in lecture notes, e.g.~\cite{Riccardo}. We hope that the reader will enjoy our presentation of this ancient topic, though, and that by taking one step back to quantum mechanics, we will be able to take two steps forward in the near future and provide incontrovertible solutions to problems like 1 and 2 above. In quantum mechanics we have the luxury of the WKB approximation, which provides us with an alternative route to the answers. Since every child knows how to use it, we will simply quote the result\footnote{Obtained by us, nota bene, in the privacy of our offices after weeks of failure!} and show that it agrees with the path integral method.

\section{Semiclassical time evolution from the path integral} \label{mainsec}
\noindent In the quantum mechanics of a particle of mass $m$ subject to a potential $V$ on the real line, an initial state with position space representation $\psi(x_0,t=0)$ evolves in time by \cite{feynman2010quantum}
\begin{align}
    \psi(x,t) &= \int_\mathbb{R} \di x_0 \overset{y(t) = x}{\underset{y(0) = x_0}{\int}} [\di y(t')] \exp \left( \frac{i}{\hbar} S[y] \right) \psi(x_0,0) \,, \label{mainPI}
\end{align}
where
\begin{equation} \label{actionfunctional}
    S[y] = \int_0^t \di t' \, \left( \frac{m}{2} \dot{y}^2 - V(y) \right) \,,
\end{equation}
and the integral in \eqref{mainPI} is over real-valued paths $y(t')$. For semiclassical states,
\begin{equation} \label{semiclassicalinitialstate}
    \psi(x_0,0) = g(x_0) \exp \left( - \frac{f(x_0)}{\hbar} \right) \,,
\end{equation}
we expect to be able to approximate \eqref{mainPI} in the limit $\hbar \to 0$ by a collection of saddle points. Physically, variations of $S$ and $f$ over the characteristic length and time scales in the problem of interest must be large compared to $\hbar$. Since the integrand involves a complex phase factor, the relevant saddle points may be complex, and we will assume $V,g$ and $f$ are analytic functions in sufficiently large regions of the complex plane to accommodate them. To highlight this fact, we will change notation $(x_0, y(t')) \to (z_0, z(t'))$ to denote complex paths.

In the method of steepest descent, what is in the exponent must be extremized with respect to all integration variables. Varying with respect to the path $z(t')$ puts us on a classical trajectory satisfying
\begin{equation} \label{classicalEOM}
    m \ddot{z} = -V'(z) \,,
\end{equation}
while varying with respect to the initial position $z_0$ sets an initial condition (e.g.\ \cite{Klauder1978,Turok:2013dfa}),
\begin{equation} \label{initialv}
    m \dot{z}(0) = i f'(z_0) \,.
\end{equation}
So we must search for solutions $\bar{z}(t')$, $t' \in [0,t]$, to \eqref{classicalEOM}, which have $\bar{z}(0) = \bar{z}_0$ for an unknown $\bar{z}_0 \in \mathbb{C}$ and initial momentum \eqref{initialv} determined by $\bar{z}_0$. This $\bar{z}_0$ should be chosen such that at the final time $t$ we have $\bar{z}(t) = x$, where $(x,t) \in \mathbb{R}^2$ are the given arguments of the wave function \eqref{mainPI}. Since $\bar{z}(t') \in \mathbb{C}$ this is one complex boundary condition, and because we have one complex parameter $\bar{z}_0$ to tune, we generally expect a discrete set of solutions.

To calculate the possible contribution to $\psi(x,t)$ from the vicinity of a saddle $(\bar{z}_0,\bar{z})$, we expand the exponent in \eqref{mainPI} to second order in the perturbations $z = \bar{z} + \delta z, z_0 = \bar{z}_0 + \delta z_0$:
\begin{equation} \label{onshell2ndorder}
    i S[z] - f(z_0) = i S[\bar{z}] - f(\bar{z}_0) - \frac{i}{2} \int_0^t \di t' \, \delta z (t') \left( m \partial_{t'}^2 + V''(\bar{z}) \right) \delta z(t') + \frac{i m}{2} \left[ \delta\dot{z} \delta z \right]^t_0 - \frac{f''(\bar{z}_0)}{2} \delta z_0^2 + \mathcal{O}(\delta^3) \,.
\end{equation}
The perturbations $\delta z(t'), t' \in [0,t]$, satisfy
\begin{equation}
    \delta z(0) = \delta z_0 \,, \quad \delta z(t) = 0 \,.
\end{equation}
To proceed, we define the fluctuation operator
\begin{equation} \label{fluctuationoperator}
    \mathcal{F} = -m \partial_{t'}^2 - V''(\bar{z})
\end{equation}
acting on functions $w(t'), t' \in [0,t]$, with $w(0) = w(t) = 0$. For real-valued $\bar{z}$, $\mathcal{F}$ is Hermitian on this space for the inner product
\begin{equation}
    \langle w, v \rangle = \int_0^t \di t' \, w(t')^* v(t') \,,
\end{equation}
so its eigenfunctions $w_n$ with eigenvalues $\lambda_n$, $n \in \mathbb{N}$, can be normalized to form an orthonormal basis. We will expand the perturbation $\delta z$ as
\begin{equation} \label{deltayexpansion}
    \delta z(t') = \alpha(t') \delta z_0 + w(t') \,,
\end{equation}
where $\alpha$ is the function satisfying $\mathcal{F}\alpha = 0$ with boundary conditions $\alpha(0) = 1, \alpha(t) = 0$ and $w$ is an arbitrary function with $w(0) = w(t) = 0$ that we expand as
\begin{equation}
    w(t') = \sum_{n=0}^\infty c_n w_n(t') \,, \quad \langle w_n, w_m \rangle = \delta_{nm} \,,
\end{equation}
for coefficients $c_n \in \mathbb{R}$. Notice that neither $\alpha$ nor the $w_n$ depend on $\delta z_0$; they depend on the background $(\bar{z}_0,\bar{z})$ only. Plugging \eqref{deltayexpansion} into \eqref{onshell2ndorder} we find
\begin{align}
    i S[z] - f(z_0) &= i S[\bar{z}] - f(\bar{z}_0) + \frac{i}{2} \sum_{n=0}^\infty \lambda_n c_n^2 - \frac{1}{2} \left[ f''(\bar{z}_0) + i m \dot{\alpha}(0) \right] \delta z_0^2 + \cdots \,.
\end{align}
Therefore, from the vicinity of this saddle, defining the integration measure
\begin{equation} \label{PImeasure}
    [\di y(t')] = \mathcal{N}_P \prod_{n=0}^\infty \di c_n
\end{equation}
up to a constant normalization factor $\mathcal{N}_P$, and integrating the $c_n$'s over $\mathbb{R}(1+i\varepsilon)$, we find
\begin{align}
    \psi(x,t) &\supset g(\bar{z}_0) \exp \left( \frac{i}{\hbar} S[\bar{z}] - \frac{f(\bar{z}_0)}{\hbar} \right) \mathcal{N}_P \sqrt{\frac{2 \pi \hbar}{f''(\bar{z}_0) + i m \dot{\alpha}(0)}} \frac{1}{\sqrt{i \operatorname{det}_D \mathcal{F}}} \left[ 1 + \mathcal{O}(\hbar) \right] ~~\text{as } \hbar \to 0 \,,
\end{align}
where $\operatorname{det}_D$ stands for the determinant with Dirichlet boundary conditions on $[0,t]$. We can compute this with the Gelfand--Yaglom method \cite{Gelfand:1959nq}, which gives
\begin{align} \label{1loopformula}
    \operatorname{det}_D \mathcal{F} = \frac{2 \pi \hbar}{m} \, \beta(t) \,,
\end{align}
where $\beta(t')$ is defined by
\begin{equation} \label{vEq}
    \mathcal{F} \beta = 0 \,, \quad \beta(0) = 0, \, \dot{\beta}(0) = 1 \,.
\end{equation}

Recall
\begin{equation} \label{uEq}
    \mathcal{F} \alpha = 0 \,, \quad \alpha(0) = 1, \, \alpha(t) = 0 \,.
\end{equation}
With this, the result can be written compactly as
\begin{align} \label{masterformula}
    \boxed{
    \begin{aligned}
        \psi(x,t) &\supset \frac{\mathcal{N}_P}{\sqrt{\gamma(t)}} \, g(\bar{z}_0) \exp \left( \frac{i}{\hbar} S[\bar{z}] - \frac{f(\bar{z}_0)}{\hbar} \right) \left[ 1 + \mathcal{O}(\hbar) \right] \quad \text{as } \hbar \to 0 \,, \\
    \text{where }& \quad    \mathcal{F} \gamma = 0 \,, \quad \gamma(0) = 1 \,, \quad m \dot{\gamma}(0) = i f''(\bar{z}_0) \,.
    \end{aligned}
    }
\end{align}
In this formula the $(x,t)$-independent normalization $\mathcal{N}_P$ has not been specified. For a general contributing saddle, unfortunately, we do not know how to determine it. For the special saddle that is continuously connected to the identity, i.e.\ the one with $\bar{z}_0 \to x$ as $t \to 0$, we have $\mathcal{N}_P = 1$ because the initial state \eqref{semiclassicalinitialstate} must be reproduced in this limit. Further, the above derivation assumed that $\bar{z}$ is real-valued, but one can show explicitly that \eqref{masterformula} solves the Schr\"odinger equation to next-to-leading order as $\hbar \to 0$ also for complex-valued $\bar{z}$, so that the formula goes through also for this case. Finally, we comment below on when saddle points indeed contribute to $\psi(x,t)$.

The generalization of \eqref{masterformula} to $d$ dimensions is straightforward. The background $(\bar{\boldsymbol{z}}_0,\bar{\boldsymbol{z}})$ solves $m \ddot{\bar{\boldsymbol{z}}} = - \nabla V(\bar{\boldsymbol{z}})$ with $m \dot{\bar{\boldsymbol{z}}}(0) = i \nabla f(\bar{\boldsymbol{z}}_0)$, and in the prefactor $\gamma(t)$ becomes $\det \gamma_{ij}(t)$ where $\sum_{j=1}^d \mathcal{F}_{ij} \gamma_{jk}(t') = 0, \gamma_{kl}(0) = \delta_{kl}, m \dot{\gamma}_{kl}(0) = i \partial_k \partial_l f(\bar{\boldsymbol{z}}_0)$ and $\mathcal{F}_{ij} = -m \delta_{ij} \partial_{t'}^2 - \partial_i \partial_j V(\bar{\boldsymbol{z}})$.

\paragraph{A word about determining the contributing saddles in the first place}
So far, we have discussed the contributions of saddle points to $\psi(x,t)$, assuming they contribute in the first place. Note that for non-quadratic systems the set of solutions to the classical boundary value problem \eqref{classicalEOM}--\eqref{initialv} is generally infinite. As for finite-dimensional integrals, however, generically not all saddle points will contribute to the semiclassical expansion of the integral. Instead, only those whose steepest \emph{ascent} surfaces intersect the original integration surface (here: real-valued functions on $[0,t]$ that attain the value $x$ at $t$) contribute; see \cite{Witten:2010cx} and references therein. Because the steepest ascent surface associated to a saddle $(\bar{z}_0,\bar{z})$ and the original integration surface both have half of the dimension of the total complexified space (complex-valued functions on $[0,t]$ that attain the value $x$ at $t$), they will generally intersect at isolated \emph{points} $(r_0,r)$, if at all. For our integral \eqref{mainPI}, concretely, we ask (see also \cite{Tanizaki:2014xba,ai2019functional}):

\vspace*{1\baselineskip}
\begingroup
\itshape
\noindent Does there exist a real-valued function $r(t'), t' \in [0,t]$, with $r(t) = x$ such that the solution $z(t';\lambda)$ to the downward flow
\begin{align} \label{downwardflowEq}
    \partial_\lambda z = -i \left( m \ddot{z}^* + V'(z^*) \right)
\end{align}
with initial condition $z(t';0) = r(t')$ for all $t' \in [0,t]$ and boundary conditions
\begin{equation}
    \partial_\lambda z(0;\lambda) = -i m \dot{z}(0;\lambda)^* + f'(z(0;\lambda))^*  \,, \quad z(t;\lambda) = x \quad \text{for all } \lambda \,,
\end{equation}
has the property that it flows to the saddle,
\begin{equation} \label{infinitycondition}
    \lim_{\lambda \to \infty} z(t';\lambda) = \bar{z}(t') \quad \text{for all } t' \in [0,t] \,?
\end{equation}
\endgroup

\noindent By analogy with finite-dimensional integrals, if the answer is yes then we expect $(\bar{z}_0,\bar{z})$ to contribute, otherwise it should not. Notice that the exponent in \eqref{mainPI} has a constant imaginary part and a monotonically decreasing real part along the flow:
\begin{align}
    \frac{\di}{\di \lambda} \left( i S[z] - f(z_0) \right) &= i m \left[ \dot{z} \, \partial_\lambda z \right]^t_0 + i \int_0^t \di t' \left( - m \ddot{z} - V'(z) \right) \times (-im \ddot{z}^* -i V'(z^*) ) - f'(z_0) \partial_\lambda z_0 \notag \\
    &= - \left| i m \dot{z}_0 + f'(z_0) \right|^2 - \int_0^t \di t' \left| m \ddot{z} + V'(z) \right|^2 \,,
\end{align}
where the subscript 0 denotes evaluation at $t' = 0$. So, two rather weak constraints on the starting point $r(t')$ of the correct flow \textdash if it exists \textdash are
\begin{align}
    - \Re f(r_0) &\geq \Re \left( i S[\bar{z}] - f(\bar{z}_0) \right) \,, \label{simpler0condition} \\
    S[r] - \Im f(r_0) &= \Im \left( i S[\bar{z}] - f(\bar{z}_0) \right) \,.
\end{align}
The flow gradually slows to zero only if a saddle point is approached.

Unfortunately we have not found an efficient way to solve \eqref{downwardflowEq}--\eqref{infinitycondition} for theories of the form \eqref{mainPI}, and leave it as an open problem.\footnote{Some obvious constraints on contributing saddles include 1) at early enough times, saddles which are naively more dominant than the special saddle that is continuously connected to the identity cannot be relevant, and 2) saddles can only become dominant as $t$ crosses $t_*$ if they were contributing subdominantly for $t < t_*$. For recent progress on this question see \cite{Shoji:2025riv}.} Instead, we will solve it implicitly in the examples we discuss below by identifying a subset of all saddle points and showing that these yield the correct result (known independently, via the WKB method or numerics). In \S\ref{numericssec} we propose a concrete, non-trivial example to which a path integral enthusiast could attempt to apply the method of \cite{Witten:2010cx} to determine the set of relevant saddles, and their normalizations $\mathcal{N}_P$, from first principles.

\section{Examples} \label{examplessec}
\noindent In the rest of this paper we will illustrate \eqref{masterformula} in a series of pedagogical problems in one-dimensional quantum mechanics. Considering that the career of a young theoretical physicist consists of treating the harmonic oscillator in ever-increasing levels of abstraction, we will start there.

\subsection{Harmonic oscillator: coherent state evolution} \label{HOsec}
\noindent Consider the quadratic potential
\begin{equation}
    V(x) = \frac{m \omega^2}{2} x^2 \,,
\end{equation}
with initial state
\begin{equation}
    \psi(x_0,0) = \mathcal{N}_0 \exp \left( - \frac{(x_0-a)^2}{2d^2} \right) \,, \quad \mathcal{N}_0^2 = \frac{1}{d \sqrt{\pi}} \,, \quad d^2 = \frac{\hbar}{m \omega} \,.
\end{equation}
This is a coherent state: its time evolution is
\begin{equation} \label{coherentstate}
    \psi(x,t) = \mathcal{N}_0 \, e^{-i \omega t/2} \exp \left( - \frac{(x-a \cos \omega t)^2}{2d^2} + i\, \frac{a \sin \omega t \left( a \cos \omega t - 2 x \right)}{2 d^2} \right) \,.
\end{equation}

To reproduce this answer with the method of \S\ref{mainsec} we must solve $\ddot{z} = -\omega^2 z$ with initial condition $\dot{z}(0) = i \omega(z_0-a)$. The solution is an ellipse centered at the origin of the complex plane,
\begin{equation} \label{ytpx0}
    \bar{z}(t') = \bar{z}_0 \cos(\omega t') + i(\bar{z}_0-a) \sin(\omega t') \,.
\end{equation}
Then we can determine $\bar{z}_0$ from the final boundary condition $\bar{z}(t) = x$, giving
\begin{equation}
    \bar{z}_0 = e^{-i \omega t} \left( x + ia \sin (\omega t) \right) \,.
\end{equation}
Calculating the on-shell exponent in \eqref{masterformula}, it is seen to match \eqref{coherentstate}. For the prefactor, we solve
\begin{align}
    \left( \partial_{t'}^2 + \omega^2 \right) \gamma = 0 \,, \quad \gamma(0) = 1 \,, \quad \dot{\gamma}(0) = i \omega \,,
\end{align}
to find
\begin{equation}
    \gamma(t') = e^{i \omega t'} \,.
\end{equation}
Evaluating at $t' = t$ and plugging into \eqref{masterformula}, setting $\mathcal{N}_P = 1$ to match the initial state, \eqref{coherentstate} is reproduced.

\subsection{Under-barrier transmission} \label{underbarriersec}
In this section, we study scattering by computing the time evolution of semiclassical energy eigenstates with energy $E < \max_{\mathbb{R}} V$. The dynamics of such states is completely captured by their transmission coefficient, through which one can subsequently reconstruct the dynamics of any localized wave packet.

\begin{figure}[!ht]
\centering
\includegraphics[width=285pt]{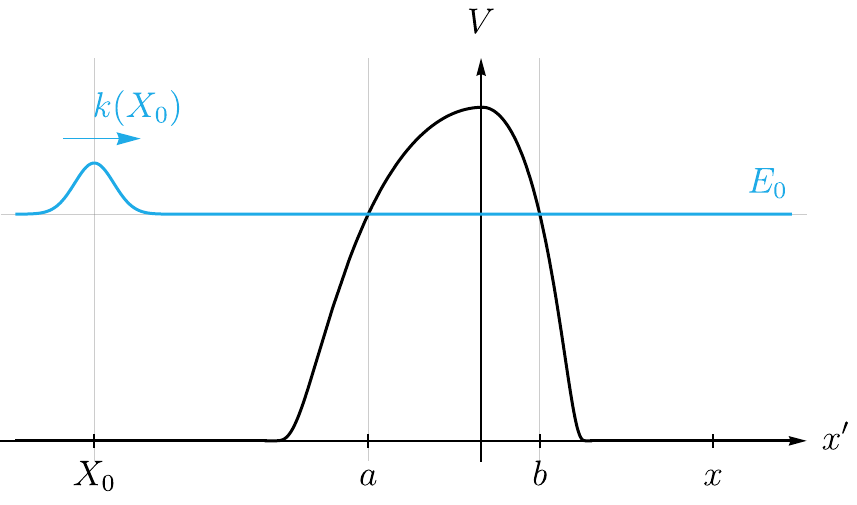}
\caption{A semiclassical wave with energy $E_0 < \max_\mathbb{R} V$ incident upon a potential barrier from the left. We are interested in computing the semiclassical wave function $\psi(x,t)$ at locations $x$ to the right of the barrier for times $t$ when the amplitude at $x$ is near its maximum.}
\label{under_barrier_setup}
\end{figure}
Consider an initial state prepared as a right-moving WKB wave at energy $E$:
\begin{equation} \label{initialstatebis}
    \psi_E(x_0,t=0) = \frac{\mathcal{N}_0}{\sqrt{k(x_0)}} \exp \left( \frac{i}{\hbar} \int_{X_0}^{x_0} k \right) \left[ 1 + \mathcal{O}(\hbar) \right] \,, \quad k(x) = \sqrt{2m \left( E - V(x) \right)} \,,
\end{equation}
to the left of a feature $V$, where we used the notation $\int f \equiv \int \di x' \, f(x')$. Here $X_0$ is an arbitrary reference point located in the free region to the left of the barrier. In this section we calculate the shape of the transmitted wave function. Our setup is summarized in Fig.~\ref{under_barrier_setup}. The classical turning points, i.e.\ the solutions to $V = E$, are labeled by $a$ and $b$.

\paragraph{WKB result}
In the WKB calculation, we assume $X_0 < a$ and $x > b$ are located in regions where the WKB approximation is valid, i.e.\ $\hbar \abs{k'} \ll k^2$. One finds
\begin{equation} \label{WKBresult1}
    \psi_E(x,t) = \frac{\mathcal{N}_0}{\sqrt{k(x)}} \exp \left[ \frac{i}{\hbar} \left( \int_{X_0}^a + \int_b^x \right) k \right] \, e^{-S/\hbar} e^{-i E t/\hbar} \left[ 1 + \mathcal{O}(\hbar) \right] \,,
\end{equation}
where
\begin{equation} \label{S0kappadef}
    S = \int_a^b \kappa \,, \quad \kappa(x) = \sqrt{2m \left( V(x) - E \right)} \,.
\end{equation}

\begin{figure}[!ht]
\centering
\includegraphics[width=300pt]{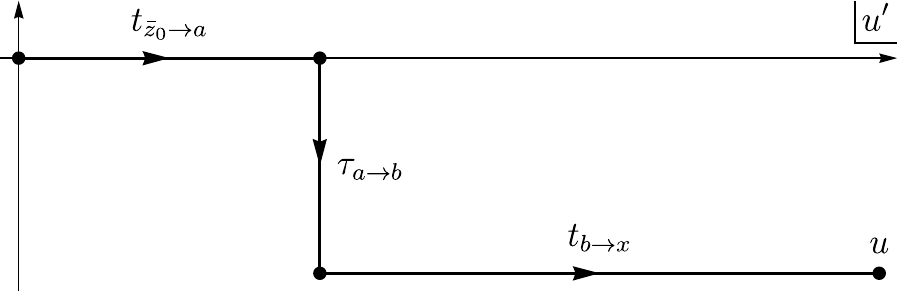}
\caption{The complexified time contour we use in our solution.}
\label{complextcontour}
\end{figure}

\paragraph{Path integral computation}
With the initial state defined by its momentum $k$, the saddle point equations and boundary condition are
\begin{equation} \label{EOMfullsaddle}
    m\ddot{\bar{z}}(t') = -V'(\bar{z}(t')) \,, \quad m \dot{\bar{z}}(0) = k(\bar{z}_0) \,, \quad \bar{z}(0) = \bar{z}_0 \,, \quad \bar{z}(t) = x \,.
\end{equation}
Of course, for a starting position $\bar{z}_0<a$ on the real line and with $t'$ kept on the real axis, the particle never reaches $x$ on the other side of the barrier where we want to evaluate the wave function. To circumvent this problem, we complexify $t \to u$ and propose instead to solve this boundary value problem on a \emph{half-line in the complex $u$-plane}, for a given real $x$ to the right of the feature. That is, we solve
\begin{equation}
    m\ddot{\bar{z}}(u') = -V'(\bar{z}(u')) \,, \quad m \dot{\bar{z}}(0) = k(\bar{z}_0) \,, \quad \bar{z}(0) = \bar{z}_0 \,, \quad \bar{z}(u) = x \,, \label{EOMfullsaddletstar}
\end{equation}
where
\begin{equation}
    u \in I \subset \mathbb{C} \,, \label{tstarregion}
\end{equation}
and $I$ will be specified shortly. The idea is to solve the equations of motion in this region of the complex plane, calculate its contribution to $\psi_E(x,u)$ according to \eqref{masterformula} by expressing the result in terms of the variable complex parameter $u$, and finally analytically continue $u\to t$.\footnote{In other words, we want to find the functions of $t \in \mathbb{R}$ for given $x \in \mathbb{R}$ appearing in the prefactor and exponent in \eqref{masterformula}, but, instead, we calculate them on a half-line in the complex $u$-plane, assume analyticity in an appropriate region, and then uniquely analytically continue $u \to t$.}
The solution to \eqref{EOMfullsaddletstar} we will consider starts at a real position $\bar{z}_0 \in \mathbb{R}$ to the left of the barrier, with a real positive initial momentum and energy $E$.
By carefully choosing the complex time contour, the particle's spatial trajectory $\bar{z}(u')$ remains real-valued throughout the entire journey. The contour consists of three segments in the complex $u'$-plane, depicted in Fig.~\ref{complextcontour}:
\begin{enumerate}
    \item Real-time approach ($0 \to t_{\bar{z}_0 \to a}$): the particle travels on the real axis with positive momentum until it comes to a halt at the WKB turning point $a$. The elapsed time is
    \begin{equation}
        t_{\bar{z}_0 \to a} \equiv \int_{\bar{z}_0}^a \frac{m}{k}
    \end{equation}

    \item Imaginary transmission ($t_{\bar{z}_0 \to a} \to t_{\bar{z}_0 \to a} - i \tau_{a \to b}$): setting $u' = t_{\bar{z}_0 \to a} - i \tau'$, with $\tau' > 0$, the particle evolves in the negative imaginary time direction. Along this segment, the particle obeys
    \begin{equation}
        \frac{m}{2} (\partial_{\tau'} \bar{z}(\tau'))^2 - V(\bar{z}(\tau')) = -E \,.
    \end{equation}
    The solution behaves as a classical particle in the inverted potential $-V$ at energy $-E$. It remains real-valued and reaches the turning point $b$ after a negative imaginary time duration
    \begin{equation}
        \tau_{a \to b} \equiv \int_a^b \frac{m}{\kappa} \,.
    \end{equation}
    \item Real-time escape ($t_{\bar{z}_0 \to a} - i \tau_{a \to b} \to u$): we transition back to real time, and the particle rolls away from the barrier, ending at $x$ after another real time duration $t_{b \to x}$.
\end{enumerate}

Summing the durations of these three segments, the final complex time $u$ is given by
\begin{equation}\label{z0relation}
    u = t_{\bar{z}_0 \to a} - i \tau_{a \to b} + t_{b \to x} \,.
\end{equation}
Since $\bar{z}_0 < a$, the domain $I$ in \eqref{tstarregion} is
\begin{equation}
    I = \{ t - i \tau_{a \to b} \, | \, t > t_{b \to x} \} \,.
\end{equation}
Turning to the calculation of the contribution to $\psi_E(x,u)$ from this saddle according to \eqref{masterformula}, we find
\begin{equation}
    i S[\bar{z}] = -i E u + i \int_{\bar{z}_0}^a k - \int_a^b \kappa + i \int_b^x k \,.
\end{equation}
Therefore, identifying $f(\bar{z}_0) = -i \int_{X_0}^{\bar{z}_0} k$ from the initial state, the on-shell exponent reads
\begin{equation}
    i S[\bar{z}] - f(\bar{z}_0) = -i E u+ i \left( \int_{X_0}^a + \int_b^x \right) k - \int_a^b \kappa \,.
\end{equation}
The prefactor $\gamma$ in \eqref{masterformula} satisfies
\begin{equation}
    \left( m \partial_{u'}^2 + V''(\bar{z}(u')) \right) \gamma = 0 \,, \quad \gamma(0) = 1 \,, \quad \dot{\gamma}(0) = \frac{k'(\bar{z}_0)}{m} \,.
\end{equation}
The solution is
\begin{equation}
    \gamma(u') = \frac{\dot{\bar{z}}(u')}{\dot{\bar{z}}(0)} \,.
\end{equation}
At time $u$, this evaluates to $k(x)/k(\bar{z}_0)$. Putting everything together in \eqref{masterformula} and using $g(\bar{z}_0) = \mathcal{N}_0/\sqrt{k(\bar{z}_0)}$, we find
\begin{align}
    \psi_E(x,u) &= \frac{\mathcal{N}_0}{\sqrt{k(x)}} \exp \left[ \frac{i}{\hbar} \left( \int_{X_0}^a + \int_b^x \right) k \right] \, e^{-S/\hbar} e^{-i E u/\hbar} \left[ 1 + \mathcal{O}(\hbar) \right] \,.
\end{align}
Here we have set the normalization constant $\mathcal{N}_P = 1$. Regarding this constant, in our procedure one could imagine increasing $x$ from near $X_0$ at early times, moving $t$ along with it on the contour shown in Fig.~\ref{complextcontour} until the point $x$ of interest to the right of the barrier is reached. In this way, our solution is continuously connected to the identity and $\mathcal{N}_P = 1$.\footnote{We wouldn't attempt to make this argument rigorous in front of our worst enemy, however.} Finally, we send $u \to t$, and assume we have found the leading saddle.
The result matches the WKB answer \eqref{WKBresult1}.

Notice that, for analytic barriers, small deviations of the strict taxicab contour shown in Fig.~\ref{complextcontour} that preserve the imaginary time displacement $-i \tau_{a \to b}$, would also end up on the other side of the barrier at time $u$. The solution $\bar{z}(u')$ would make a small incursion into the complex plane in this representation of it. We only sum over one such representative in \eqref{masterformula}.

\paragraph{A word about classical wave packets}
We could also study the scattering of a localized particle at energy $E_0$, such as a Gaussian wave packet starting to the left of the feature:
\begin{equation}\label{eq:gaussian_wavepacket1}
    \psi(x_0,0) = \mathcal{N}_\alpha \exp \left( - \frac{\alpha}{2\hbar} (x_0-X_0)^2 + \frac{i}{\hbar} \int_{X_0}^{x_0} k_0 \right) \,, \quad \mathcal{N}_\alpha^4 = \frac{\alpha}{\pi \hbar} \,, \quad k_0 = k(E_0) \,.
\end{equation}
Provided $k_0$ and $V$ vary slowly over the characteristic width of the wave packet,\footnote{Specifically, $\abs{k_0'(X_0)} \sqrt{\hbar/\alpha} \ll \sqrt{\alpha \hbar} \ll k_0(X_0)$.} its spatial variance $\sigma_x^2 \approx \hbar/2\alpha$ and momentum variance $\sigma_p^2 \approx \alpha \hbar /2$ are both proportional to $\hbar$, ensuring that the wave packet is tightly localized in phase space in the semiclassical limit.
One could compute the time evolution of this state by decomposing it into the energy eigenstates we computed above, but it is instructive to think about how the path integral computation for this packet proceeds. The initial momentum boundary condition now acquires an imaginary shift: $m \dot{\bar{z}}(0) = k(\bar{z}_0) + i \alpha(\bar{z}_0 - X_0)$. This means that for general $\alpha$, we should consider complex trajectories. However, to determine the wave function to first order as $\alpha \to 0$, corresponding to a wave packet that is strongly peaked in energy around $E_0$, it is sufficient to solve the equations of motion with $\alpha=0$.\footnote{The real part of $\log\psi$ is already $\mathcal{O}(\alpha)$ so a linear change in $\alpha$ to the background would not affect it, and the modifications to the on-shell action and on-shell momentum term in \eqref{eq:gaussian_wavepacket1} are $\mathcal{O}(\alpha^2)$ because the $\alpha = 0$ equations of motion are solved. At order $\alpha^2$ one uncovers that the transmitted pulse travels \emph{faster} than the incoming one \textdash because higher-energy modes are transmitted more easily \textdash and at order $\alpha^3$ one starts to see non-Gaussianity.} This yields
\begin{align}
    \psi(x,t) &= \sqrt{\frac{k_0(\bar{z}_0)}{k_0(x)}} \, \mathcal{N}_\alpha \exp \left[ -\frac{\alpha}{2 \hbar} \left( \bar{z}_0 - X_0 \right)^2 + \frac{i}{\hbar} \left( \int_{X_0}^a + \int_b^x \right) k_0 \right] \, e^{-S_0/\hbar} e^{-i E_0 t/\hbar} \left[ 1 + \mathcal{O}(\hbar) \right] \,,
\end{align}
characterized by the exponentially suppressed transmission coefficient $e^{-S_0/\hbar}$ where $S_0 \equiv S(E_0)$. The location of the center of the transmitted wave packet emerges by evaluating the initial saddle point position $\bar{z}_0$ associated with the unperturbed $\alpha = 0$ solution as a function of $x$ and $t$ via \eqref{z0relation}, after sending $u \to t$. To make this more explicit we will assume the particle starts in a free region to the left of the feature where $V \to 0$, with momentum $k_0(\bar{z}_0) = P_0$. We find $\bar{z}_0$ by inverting \eqref{z0relation},
\begin{align}
    \bar{z}_0= x - \frac{P_0}{m} \left( t - t_{\textsf{offset}} + i \tau_{a \to b} \right) \,,
\end{align}
where
\begin{align}
    t_{\textsf{offset}} &= t_{X_0 \to a} + t_{b \to X_1} - \frac{m}{P_0}(X_1-X_0)
\end{align}
is the semiclassical time delay (or advance) \textdash representing the difference between the time it takes to traverse the classical regions near the potential feature and the time a free particle would take.\footnote{The point $X_1$ we introduced here is arbitrary, provided it lies in the free region to the right of the barrier.}

\paragraph{A word about the sign}
One may wonder how the sign of the prefactor $1/\sqrt{\gamma(t)}$ in \eqref{masterformula} is determined, especially if the connection to the identity solution at $t = 0$ is opaque as in our calculation above. We found that the prescription of going counterclockwise around WKB turning points when coming from a real-time segment, and clockwise around WKB turning points when coming from an imaginary-time segment, produces the correct sign. These prescriptions ensure the time evolution operator $e^{-iHt'}$ remains bounded as we transition between the segments of real and imaginary time evolution. In Fig.~\ref{complextcontour}, the prefactor picks up a phase $e^{-i \pi/4}$ when going from the first real-time segment to the imaginary one, and another phase $e^{i \pi/4}$ when transitioning to the second real-time segment. Therefore the sign is positive in this example.

\paragraph{A word about complex solutions in real time vs.\ real solutions in complex time}
The example in \S\ref{HOsec} proceeded by keeping $t' \in [0,t]$ real; no analytic continuation in $t$ was performed there. One could call this the \enquote{direct} method of evaluating \eqref{masterformula}: time is kept real throughout, and, in general, the solutions one finds are complex-valued. Mathematically, this approach is straightforward. However, in certain situations, the \enquote{indirect} method we employed above \textdash that is, complexifying time while keeping the solution real-valued throughout, and analytically continuing time back to the real axis at the end of the calculation \textdash may be more insightful. This alternative approach seems to us particularly useful to study quantum tunneling.

A practical issue with the direct method is that it appears one must be able to explicitly solve the equations of motion to proceed analytically, which is only possible in special cases. Moreover, the solution is sensitive to the analytically continued potential $V(z)$ in the complex plane. For the problem of under-barrier transmission we treated here, and the decay of a metastable state that we will discuss in \S\ref{decaysec}, the result is independent of the properties of $V$ in the complex plane to all orders in perturbation theory in $\hbar$. This means that \textdash for these problems \textdash one could have added a contribution to $V$ which has a negligible effect on the answer, but which would have a disastrous effect on the direct method, making it practically infeasible. The indirect method avoids this issue.

Some quantities of interest, however, \emph{are} sensitive to the complex-analytic properties of $V$. We discuss an example in the next section. We will use a complex solution in complex time to understand it.

\subsection{Over-barrier reflection} \label{overbarriersec}
\begin{figure}[!ht]
\centering
\includegraphics[width=330pt]{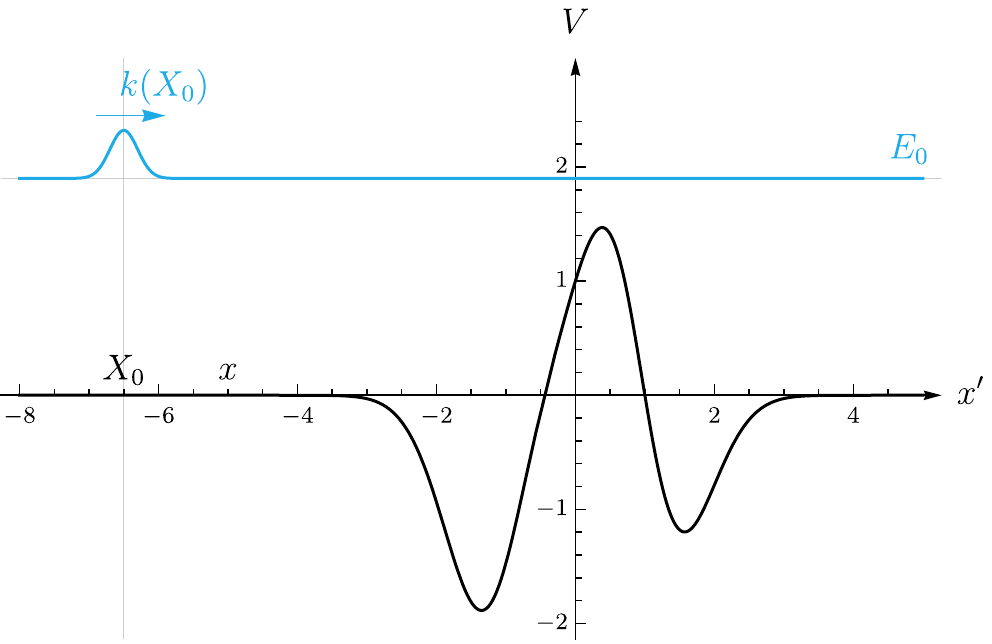}
\caption{A semiclassical wave with energy $E_0 > \max_\mathbb{R} V$ incident upon a potential barrier from the left. We want to know what the \emph{reflected} wave looks like.}
\label{over_barrier_setup}
\end{figure}

\noindent The setup in this section is identical to that of \S\ref{underbarriersec}, except that the energy $E_0$ of the incoming wave lies well \emph{above} the barrier and we are interested in computing the \emph{reflected} wave. We assume the potential $V$ is an analytic function in a large enough region of the complex plane that contains the real axis.

\paragraph{WKB result}
We first review the result of \cite{pokrovskii1961problem} (see \cite{Berry1982} for a generalization and historical review of this problem), assuming $V(\pm \infty) = 0$. Of the two energy eigenstates with energy $E =p^2/2m$, the one relevant to our problem has asymptotics
\begin{align}
    \psi_p(x) \sim \left\{
    \begin{array}{ll}
        e^{ipx/\hbar} + r(p) e^{-ipx/\hbar} & \text{as } x \to -\infty \,, \\
        t(p) e^{ipx/\hbar} & \text{as } x \to +\infty \,.
    \end{array}
\right.
\end{align}
For an analytic barrier, \cite{pokrovskii1961problem} finds
\begin{equation}
    r(p) \sim a(p) e^{b(p)/\hbar} \quad \text{as } \hbar \to 0 \,,
\end{equation}
with
\begin{equation} \label{aofpbofp}
    a(p) = -i \,, \quad b(p) = 2i X_0 p + 2i \int_{X_0}^{z_c} k \,,
\end{equation}
where $X_0$ is an arbitrary point in the free region (notice that its dependence drops out). Here $z_c \in \mathbb{C}$ is the complex WKB turning point (i.e.\ it solves $V(z_c) = E$) with the smallest positive value for $- \Re b(p)$, that is, it is the largest contribution to $r$ that is still exponentially suppressed as $\hbar \to 0$.\footnote{More precisely we have assumed $z_c$ is a \emph{linear} turning point, $V'(z_c) \neq 0$. Higher-order turning points alter the prefactor $a(p)$ \cite{Berry1982}.} The integral in \eqref{aofpbofp} should be taken along a contour that picks up no contributions from other branch points $\tilde{z}_c \neq z_c$.

\paragraph{Path integral computation}
As in \S\ref{underbarriersec} we will search for a trajectory satisfying \eqref{EOMfullsaddle} on a complex time contour. What remains to be determined is the set of values $\{ u \} \subset \mathbb{C}$ that we will analytically continue the result from. To find it, notice that although $E_0 = V(x)$ does not have real solutions, it will have complex ones; let us focus on a particular solution $z_c \in \mathbb{C}$. In the first instance, reasoning in analogy with under-barrier transmission, we would like the particle to reach this turning point and then travel backward along its original path: in this way we would reverse the particle's momentum (in the complex plane, now) and end up back on the real axis with opposite momentum \textdash something that was impossible while remaining on the real $u'$-axis. By energy conservation, the complex time $u_c$ at which the complex turning point $z_c$ is reached, is given by the contour integral
\begin{equation} \label{tstarcontour}
    u_c = \sqrt{\frac{m}{2}} \int_{\bar{z}_0}^{z_c} \di z \frac{1}{\sqrt{E_0 - V(z)}} \,.
\end{equation}
The piecewise straight lines in the complex $u'$-plane,
\begin{equation} \label{straightlinepath}
    u'\colon \quad 0 \to \Re u_c \to \Re u_c + i \Im u_c \to \Re u_c + 2 i \Im u_c \to \Re u_c + 2 i \Im u_c + t_{a \to x} = u \,,
\end{equation}
correspond to the path $z(u') \in \mathbb{C}$ which first moves toward the barrier on the real line until a specific point $a \in \mathbb{R}$, then makes a complex (generally, curved) transition, reaching the complex turning point $z_c \in \mathbb{C}$ at time $u_c = \Re u_c + i \Im u_c$, where it comes to a halt. Then the path is backtracked, reaching the real line once more at $u' = \Re u_c + 2 i \Im u_c$, and finally moving on the real line \textdash away from the barrier, now \textdash for time $t_{a \to x}$ until the point $x$ is reached. So the imaginary part of the half-line we will analytically continue from is $2 \Im u_c$.

\begin{figure}[p]
\centering
\includegraphics[width=220pt]{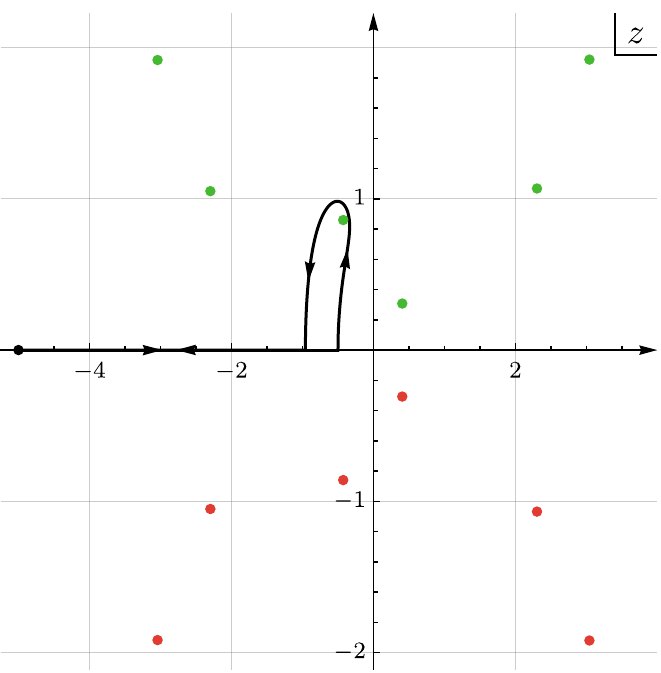}
\hfil \hfil
\includegraphics[width=220pt]{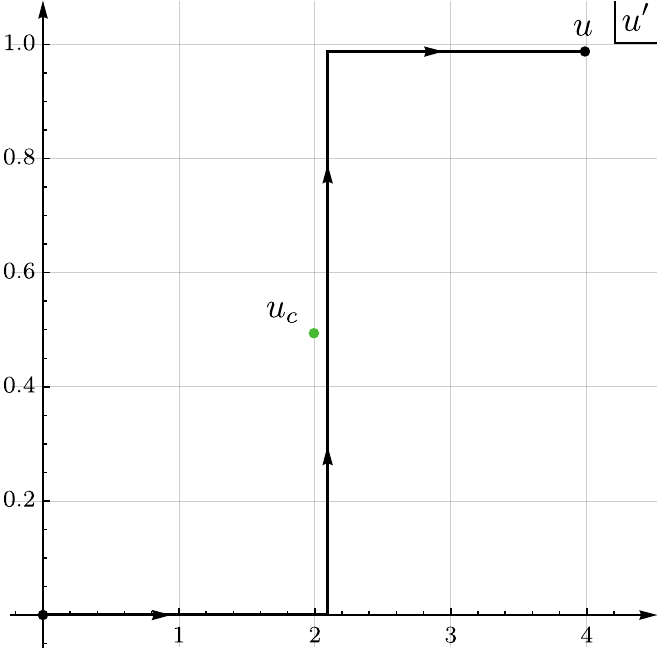}
\caption{A slight deformation of the $u'$-contour corresponding to a subleading saddle, keeping the imaginary displacement $\Delta \Im u = 2 \Im u_c$ fixed but passing $u_c$ to the \emph{right}. This causes the semiclassical particle to travel a bit farther on the real line before making the detour in the complex plane, so that the turning point $z_c \in \mathbb{C}$ is encircled in the \emph{counterclockwise} direction.}
\label{over_example3}
\end{figure}

Small deformations of the path \eqref{straightlinepath} which preserve the total imaginary time displacement $2 i \Im u_c$ and which are parallel to the real axis at early and late times, do not destroy the essential property of this solution, namely, that it starts on the real axis with momentum pointing toward the barrier and ends on the real axis with momentum pointing away from the barrier. We find that choosing a deformation which passes \emph{counterclockwise} around the turning point $z_c$ yields the correct sign for the reflected wave function (see Fig.~\ref{over_example3}).\footnote{In the $u'$-plane, for saddles with $\Im u_c < 0$ (hence $\Im z_c < 0$, which we conclude because the momentum on the real $u'$-axis is positive), this means passing to the \emph{left} of $u_c$, while for saddles with $\Im u_c > 0$ (so $\Im z_c > 0$) this means passing to the \emph{right} of $u_c$.}

\begin{figure}[p]
\centering
\includegraphics[width=220pt]{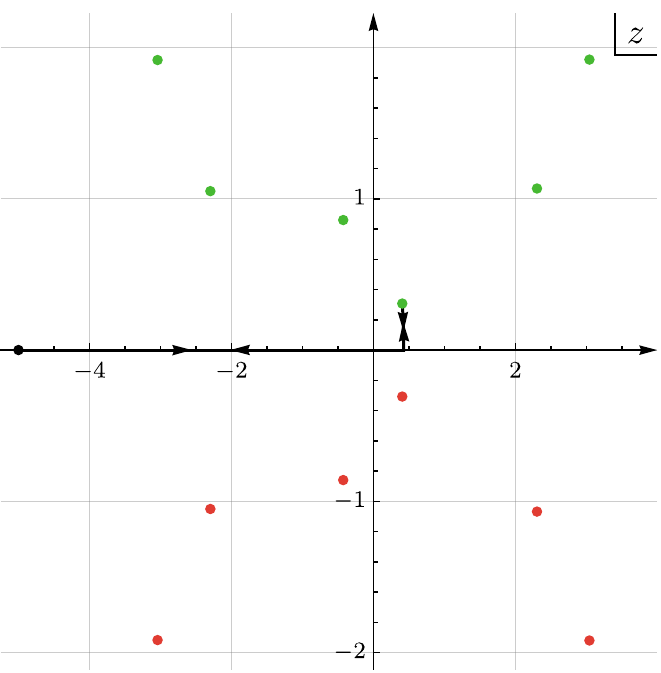}
\hfil \hfil
\includegraphics[width=220pt]{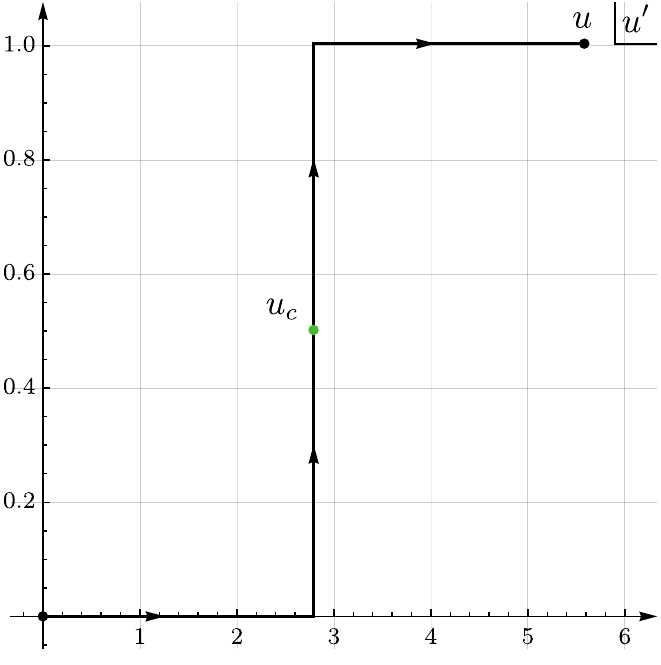}
\caption{On the left we show solutions to $V(z_c) = E_0$ (colored dots). The green points give exponentially suppressed contributions to $\psi$ while the red points would give exponentially enhanced contributions, but they must be excluded as they violate \eqref{simpler0condition}. We consider a particular solution $z_c$ (here, the green point closest to the real axis) and compute for it the complex turning point $u_c$ in \eqref{tstarcontour}; this is the point reached exactly halfway through the complex time contour on the right plot. Following the contour in \eqref{straightlinepath}, the solution reaches the turning point and then retraces its steps. For this solution $\Re \left( i S \right) = -0.46$; it is the dominant contribution to $\psi$.}
\label{over_example1}
\end{figure}

Before continuing with the calculation, we illustrate the reasoning above in Figs.~\ref{over_example1} and \ref{over_example3}, for the potential
\begin{equation} \label{scatteringV}
    V(y) = \left( 1 +2y-3y^4 \right) e^{-y^2} \,, \qquad \text{(Fig.~\ref{over_barrier_setup})}
\end{equation}
and parameters $x = -5$, $E_0 = 1.3 \times \max_\mathbb{R} V$ and with $u \in \mathbb{C}$ chosen such that $\bar{z}_0 = x$.

The calculation of the wave function proceeds as in \S\ref{underbarriersec}. For simplicity, we will assume $V \to 0$ as $y \to -\infty$ and that $X_0$, $x$ and so also $\bar{z}_0$ lie in this region. The momentum of the incoming particle is $P_0 = \sqrt{2m E_0}$. To analytically continue, we will use
\begin{align}
    u = \frac{m}{P_0} \left( 2 X_0 - x - \bar{z}_0 \right) + 2 t_{X_0 \to z_c} \,. \label{deltatdef}
\end{align}
For the on-shell exponent we find
\begin{align}
    i S[\bar{z}] - f(\bar{z}_0) &= -i E_0 u - i P_0(x + X_0) + b(P_0) \,,
\end{align}
with $b$ as in \eqref{aofpbofp}. This matches the WKB result \eqref{aofpbofp}, and by energy conservation, the prefactor is $\gamma(t)^{-1/2}=\sqrt{-1}$, where the correct sign is obtained by going counterclockwise around $z_c$.

Finally, we can apply this result to compute the reflection of the localized classical wave packet introduced in \eqref{eq:gaussian_wavepacket1}, which is strongly peaked around energy $E_0$. The path integral formulation yields an undeformed reflected Gaussian wave to leading order as $\alpha \to 0$, where the reflection coefficient is simply $r(P_0) = -i \, e^{b(P_0)/\hbar}$, and the time delay $t_\textsf{offset}$ is given by
\begin{equation}
    t_\textsf{offset} = - \frac{i m}{P_0}b'(P_0) = \frac{2m}{P_0} \left( X_0 + \sqrt{\frac{m}{2}} \int_{X_0}^{z_c} \frac{\di z}{\sqrt{E_0 - V(z)}} \right) \,.
\end{equation}

\clearpage
\subsection{Decay of a metastable state} \label{decaysec}
\begin{figure}[!ht]
\centering
\includegraphics[width=330pt]{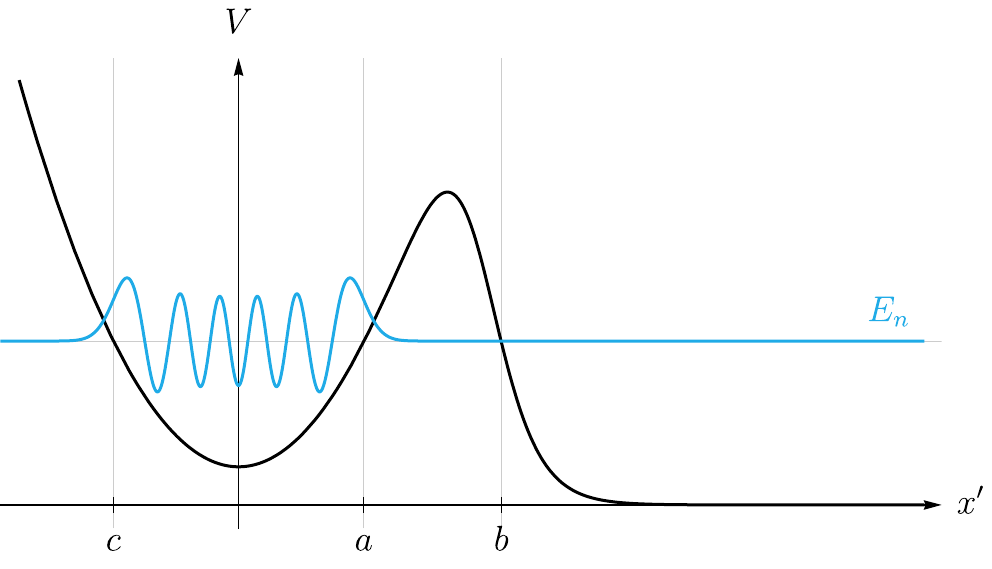}
\caption{The potential $V$ (black) has a false vacuum that is separated from a free region by a barrier. We consider an initial state (blue) that is localized in the metastable region and is largely composed of energy eigenstates with energies between the bottom of the false vacuum and the top of the barrier. We assume there are many resonant energy levels $E_n$ of the false vacuum (i.e.\ many WKB levels \eqref{WKBquant}) and that the \enquote{barrier integrals} $S_n/\hbar$ are large, with $S_n = S(E_n)$ given in \eqref{alphatheta}. Formally these requirements can be achieved by having a \enquote{classical potential} (independent of $\hbar$) and sending $\hbar \to 0$.}
\label{tunneling_setup}
\end{figure}

In this section we study the decay of a metastable state, that is, a state that is initially localized in the basin of attraction of a minimum of the potential and is mainly composed of energy eigenstates with energy below the height of the barrier. Probability slowly leaks out of the well by quantum tunneling. Our setup is summarized in Fig.~\ref{tunneling_setup}.

More precisely, the spectrum of $H$ is continuous and non-degenerate. The energy eigenfunctions with significant relative support in the metastable region are localized in energy bands of order $\Delta E_n \sim (\hbar/t_n) e^{-2S_n/\hbar}$ around energies $E_n$ that satisfy the WKB quantization condition of the metastable region. Define
\begin{equation} \label{alphatheta}
    \phi(E) = \int_{c}^{a} k \,, \quad S(E) = \int_{a}^{b} \kappa \,,
\end{equation}
with $k(x;E)$ as in \eqref{initialstatebis} and $\kappa(x;E)$ as in \eqref{S0kappadef}, and $c,a,b$ the classical turning points shown in Fig.~\ref{tunneling_setup}. The levels $E_n$ are defined by
\begin{equation} \label{WKBquant}
    \phi(E_n) = \pi \hbar \left( n + 1/2 \right) \,, \quad n \in \{ 0,1,\dots,N_\textsf{max}\} \,,
\end{equation}
$S_n \equiv S(E_n)$ and the timescale $t_n$ is the classical oscillation time at energy $E_n$ in the well,
\begin{equation}
    t_n = 2 t_{c \to a}\big|_{E=E_n}
\end{equation}
in the notation of \S\ref{underbarriersec}. The decay rate of a metastable state with predominant support in the band around $E_n$ is of order
\begin{equation} \label{Gammandef}
    \frac{\Delta E_n}{\hbar} \sim \frac{1}{t_n} e^{-2S_n/\hbar} \equiv \Gamma_n \,.
\end{equation}

In the first instance (\S\S\ref{outsidesec}--\ref{backoutsidesec}), we will assume the initial state is composed of \emph{excited} resonant states. So we will assume it is possible to decompose, to good approximation, the initial state as
\begin{equation} \label{decomppsit0}
    \psi(x_0,t=0) = \sum_{n=N_\textsf{min}}^{N_\textsf{max}} c_n \tilde{\psi}_n(x_0) \,, \quad 1 \ll N_\textsf{min} \ll N_\textsf{max} \,.
\end{equation}
The $\tilde{\psi}_n$ are defined in the region $x_0 \leq x_*$ as proportional to the true energy eigenstate at energy $E_n$, where $x_* \in [a,b]$ is a point in the forbidden region where the wave function is exponentially small. In WKB we have, for a state with $n \gg 1$ and away from the turning points,
\begin{align} \label{WKBpsin}
    \tilde{\psi}_n(x) = \mathcal{N}_n \left\{
    \begin{array}{ll}
        \displaystyle\frac{1}{\sqrt{\kappa_n}} \exp \left( - \frac{1}{\hbar} \displaystyle\int_x^{c} \kappa_n \right) & \text{for } x \leq c - \Delta_c \,, \\[2ex]
        \displaystyle\frac{2}{\sqrt{k_n}} \cos \left( \frac{1}{\hbar} \displaystyle\int_{c}^x k_n - \frac{\pi}{4} \right) & \text{for } c + \Delta_c \leq x \leq a - \Delta_a \,, \\[2ex]
        \displaystyle\frac{(-1)^n}{\sqrt{\kappa_n}} \exp \left( - \frac{1}{\hbar} \int_{a}^x \kappa_n \right)  & \text{for } a + \Delta_a \leq x \leq x_* \,,
    \end{array}
    \right.
\end{align}
where $\Delta_{c,a} \to 0$ as $\hbar \to 0$ and we have suppressed the $1 + \mathcal{O}(\hbar)$ correction. The point $x_*$ should be chosen deep in the forbidden region such that for the highest energy level $n$ in \eqref{decomppsit0} on which the initial state has significant support, $\int_{a}^{x_*} \kappa_n/\hbar \gg 1$. The main physics we will be interested in describing \textdash namely, the wave function somewhat outside the well at times $t_n \lesssim t \lesssim \mathcal{O}(1)/\Gamma_n$ (\S\ref{outsidesec} and \S\ref{backoutsidesec}) and the wave function inside the well at times $0 \leq t \lesssim \mathcal{O}(1)/\Gamma_n$ (\S\ref{insidesec}) \textdash is not sensitive to the exact placement of $x_*$ or the precise way in which the $\tilde{\psi}_n$ are defined beyond $x_*$ as long as they continue to decay.\footnote{To be complete, one could define an auxiliary potential $\tilde{V}$ that is equal to $V$ for $x \leq x_*$ but grows indefinitely beyond it, and define the $\tilde{\psi}_n$ as the (discrete, normalized) energy eigenstates of $\tilde{V}$. We leave this arbitrary machinery implicit.} At large $n$ the normalization factor in \eqref{WKBpsin} can be approximated by noting that the main contribution to $\int_\mathbb{R} \abs{\tilde{\psi}_n}^2$ comes from the classically allowed region. In this region the wave function rapidly oscillates, which approximately kills half of the integral:
\begin{equation}
    \int_{c}^{a} \di x \, \frac{4}{k_n(x)} \cos^2 \left( \frac{1}{\hbar}\displaystyle\int_{c}^x k_n - \frac{\pi}{4} \right) \approx 2 \int_{c}^{a} \frac{1}{k_n} = \frac{t_n}{m} \,,
\end{equation}
so that
\begin{equation} \label{Nnnorm}
    \mathcal{N}_n \approx \sqrt{\frac{m}{t_n}} \,.
\end{equation}
The lower bound $N_\textsf{min} \gg 1$ we introduced in \eqref{decomppsit0} is merely for convenience, so that we can use the approximate WKB form \eqref{WKBpsin} of the $\tilde{\psi}_n$. Our discussion can be extended to low-lying states, which we comment on in \S\ref{numericssec}. Finally, by linearity of the Schr\"odinger equation, we have
\begin{equation}
    \psi(x,t) = \sum_{n=N_\textsf{min}}^{N_\textsf{max}} c_n \psi_n(x,t)
\end{equation}
where $\psi_n$ is the time evolution of the initial state $\tilde{\psi}_n$. We will focus on an individual state labeled by $n$ in what follows.

\subsubsection{Wave function outside the metastable region at intermediate times} \label{outsidesec}
Here we would like to compute the wave function $\psi_n(x,t)$ for $x > b$ in the free region and for times $t_n \lesssim t \lesssim 1/\Gamma_n$. (For times $t \ll 1/\Gamma_n$ this calculation shares the essential features of the under-barrier transmission calculation of \S\ref{underbarriersec}.) The reason for the lower bound on time is that at very early times, $\psi_n(x,t)$ is sensitive to the precise form of the initial state outside the false vacuum, which we have left unspecified in our decomposition \eqref{decomppsit0}. We will return to the early-time behavior of the wave function outside the well in \S\ref{numericssec} in a specific example; it is non-universal. The reason for the upper bound on time is similar: at very late times $t \gg 1/\Gamma_n$, the presence of an (effective) lowest energy in the decomposition of $\tilde{\psi}_n$ into the true eigenstates of $H$ will be felt. This typically causes the exponential falloff in $\psi_n(x,t)$ that we will find for $t \lesssim 1/\Gamma_n$ to transition into a polynomial tail for $t \gg 1/\Gamma_n$ \cite{Khalfin}.
The description of this transition is likewise non-universal.

\paragraph{WKB result}
\begin{equation} \label{psiXToutside}
    \psi_n(x,t) = \mathcal{N}_n \frac{(-1)^n}{\sqrt{k_n(x)}} \exp \left( \frac{i}{\hbar} \int_{b}^x k_n + \frac{i \pi}{4} \right) e^{-S_n/\hbar} e^{-i E_n t/\hbar} \times e^{-\Gamma_n t/2} \left[ 1 + \mathcal{O}(\hbar) \right] \,,
\end{equation}
where, recall, $\Gamma_n$ was defined in \eqref{Gammandef}.

To reproduce this result with \eqref{masterformula}, it is instructive to separately consider the regimes $t_n \lesssim t \ll 1/\Gamma_n$, where the exponential damping factor in \eqref{psiXToutside} has not yet set in, and $t \lesssim 1/\Gamma_n$. We will address the former regime first, coming back to the latter in \S\ref{backoutsidesec}.

\paragraph{Path integral computation for $t_n \lesssim t \ll 1/\Gamma_n$}
We will assume the initial position $\bar{z}_0$ of the background lies in the classically allowed region near the false vacuum, so the initial state, from \eqref{WKBpsin}, is effectively
\begin{equation}
    \psi_n(x_0,t=0) \approx \frac{2\mathcal{N}_n}{\sqrt{k_n(x_0)}} \cos \left( \frac{1}{\hbar} \int_{c}^{x_0} k_n - \frac{\pi}{4} \right) \,.
\end{equation}
For the time-evolved state we can write, schematically,
\begin{equation} \label{psiXTPI}
    \psi_n(x,t) = \int \di x_0 \overset{y(t) = x}{\underset{y(0) = x_0}{\int}} [\di y(t')] \left( \Circled{$R$} + \Circled{$L$} \right) \,,
\end{equation}
where
\begin{equation}\label{RLdefinition}
    \Circled{$R$} = \frac{\mathcal{N}_n}{\sqrt{k_n(x_0)}} \exp \left[ \frac{i}{\hbar} \left( S[y] + \int_{c}^{x_0} k_n \right) - \frac{i\pi}{4} \right] , \quad \Circled{$L$} = \frac{\mathcal{N}_n}{\sqrt{k_n(x_0)}} \exp \left[ \frac{i}{\hbar} \left( S[y] - \int_{c}^{x_0} k_n \right) + \frac{i \pi}{4} \right] .
\end{equation}
The saddle point equations are
\begin{align} \label{saddleEOM}
    m \ddot{\bar{z}} = -V'(\bar{z}) \,, \quad m \dot{\bar{z}}(0) = \pm k_n(\bar{z}_0) \,, \quad \bar{z}(0) \equiv \bar{z}_0 \,, \quad \bar{z}(u) = x \,,
\end{align}
with $+$ for the (initially) right-moving branch \Circled{$R$} and $-$ for the (initially) left-moving one \Circled{$L$}. Both imply the conservation of energy
\begin{equation}
    \frac{m}{2} \dot{\bar{z}}^2 + V(\bar{z}) = E_n \,.
\end{equation}
The on-shell exponent $i \bar{F}/\hbar$ of such solutions is calculated as
\begin{align} \label{Fexpression}
    \bar{F}_\pm = S[\bar{z}] \pm \int_{c}^{\bar{z}_0} k_n \mp \frac{\pi \hbar}{4} = -E_n u + \int_0^{u} \di u' \, \dot{\bar{z}} (m \dot{\bar{z}}) \pm \int_{c}^{\bar{z}_0} k_n \mp \frac{\pi \hbar}{4} \,.
\end{align}
In what follows, during real time evolution we will change variables in the $u'$ integral from $u' \to \bar{z}$ using $m \dot{\bar{z}}(u') = \pm k_n(\bar{z}(u'))$, remembering that all contributions are positive. During imaginary time evolution we will use $m \partial_\tau \bar{z} = \kappa_n(\bar{z})$.

\begin{figure}[!ht]
\centering
\includegraphics[width=300pt]{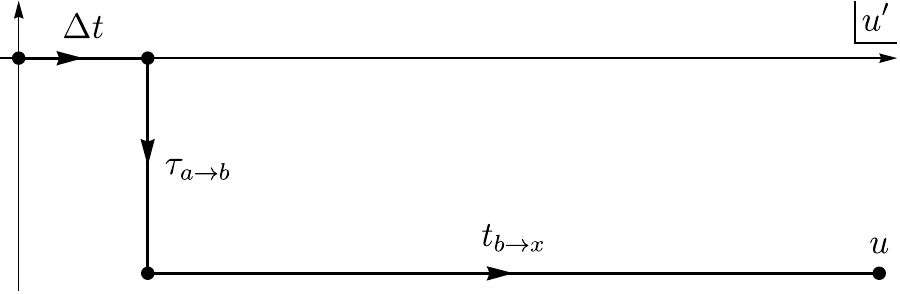}
\caption{The complex time contour and semiclassical particle trajectory used to calculate the wave function outside the well at \enquote{early} complex final times $u = \Delta t + t_{b \to x} - i \tau_{a \to b}$ with $\Delta t < t_n/2$.}
\label{singletunnelingfig}
\end{figure}

The complexified time solutions we will consider are similar to those of \S\ref{underbarriersec}. We will also analytically continue from the half-line
\begin{equation} \label{tstarhalfline}
    u \in \{ \Delta t + t_{b \to x} - i \tau_{a \to b} \, | \, \Delta t > 0 \} \,,
\end{equation}
with $a, b$ the classical turning points shown in Fig.~\ref{tunneling_setup}. For small $\Delta t < t_n/2$ in \eqref{tstarhalfline}, given our assumption that $\bar{z}_0$ lies in the classically allowed region, only the saddle point equations associated with the \Circled{$R$} branch have a solution: the particle starts with a position $\bar{z}_0$ to the left of $a$, with positive momentum. After a time $\Delta t$ we reach $a$, then we travel in the negative imaginary time direction, reaching the other side of the barrier $b$ after time $\tau_{a \to b}$. After another real time $t_{b \to x}$ we reach the final point $x$. The left-moving branch gives no contribution for these \enquote{early} complex $u$. This is illustrated in Fig.~\ref{singletunnelingfig}.

The contribution of this saddle is
\begin{align}
    \psi_n(x,u) &= \sqrt{\frac{k_n (\bar{z}_0)}{k_n(x)}} \times \frac{\mathcal{N}_n}{\sqrt{k_n(\bar{z}_0)}} e^{i \bar{F}_+/\hbar} \left[ 1 + \mathcal{O}(\hbar) \right] \,, \quad \Delta t < t_n/2 \,, \\
    \bar{F}_+ &= -E_n u + i \int_{a}^{b} \kappa_n + \int_{b}^x k_n + \int_{c}^{a} k_n - \frac{\pi \hbar}{4} \notag \\
    &= -E_n u + i \int_{a}^{b} \kappa_n + \int_{b}^x k_n + \pi \hbar n + \frac{\pi \hbar}{4} \,,
\end{align}
where the prefactor, including its sign, was computed as in \S\ref{underbarriersec}, and we used the WKB quantization condition of the $n$th level. Sending $u \to t$, this expression already matches the WKB result \eqref{psiXToutside}, without the final exponential damping factor.

\begin{figure}[!ht]
\centering
\includegraphics[width=300pt]{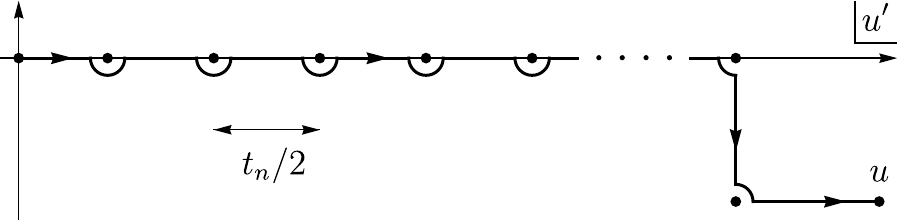}
\caption{The complex time contour we use for times $u = \Delta t + t_{b \to x} -i \tau_{a \to b}, \Delta t > 0$. On the real axis, we go around all WKB turning points (black dots) in the lower-half plane. For each, the prefactor picks up a phase $e^{-i\pi/2}$, which cancels with an opposite phase coming from the change in on-shell exponent when going from a right-moving to a left-moving solution or vice versa.}
\label{singletunnelingfig2}
\end{figure}

Now, when complex time $u$ in \eqref{tstarhalfline} is reached with $\Delta t$ crossing $t_n/2$, the right-moving solution for $\Delta t < t_n/2$ turns into a (initially) left-moving solution for $\Delta t > t_n/2$. In this regime, only the \Circled{$L$} branch contributes to $\psi$. The on-shell exponent now evaluates to
\begin{equation}
    \bar{F}_- = \bar{F}_+ + \frac{\pi \hbar}{2} \,,
\end{equation}
so the wave function picks up a phase $e^{i \pi/2}$ from this. This phase is canceled by a contribution $e^{-i\pi/2}$ coming from the prefactor $\sqrt{\dot{\bar{z}}(0)/\dot{\bar{z}}(u)}$, however, following the prescription of going counterclockwise around WKB turning points when time runs on the real axis, as discussed in \S\ref{underbarriersec}. Therefore our saddle point approximation for $\psi_n(x,u)$ is continuous as $\Delta t$ crosses $t_n/2$, and, likewise, for all transitions across $\Delta t = N t_n/2$. We illustrate this in Fig.~\ref{singletunnelingfig2}. This concludes the semiclassical path integral computation of $\psi_n(x,t)$ for $x$ outside the false vacuum for times $t_n \lesssim t \ll 1/\Gamma_n$.

\subsubsection{Wave function inside the metastable region} \label{insidesec}
We now turn to the wave function inside the false vacuum, $x \in (c, a)$ in Fig.~\ref{tunneling_setup}, for times $0 \leq t \lesssim 1/\Gamma_n$.

\paragraph{WKB result}
\begin{equation} \label{psixtint3}
    \psi_n(x,t) = \tilde{\psi}_n(x) \, e^{-i E_n t/\hbar} \times e^{-\Gamma_n t/2} \left[ 1 + \mathcal{O}(\hbar) \right] \,.
\end{equation}

\paragraph{Path integral computation}
At early times $t \ll t_n$, clearly, there is one contribution to \Circled{$R$} (with $\bar{z}_0^R < x$) and one contribution to \Circled{$L$} (with $\bar{z}_0^L > x$), which do not require complexifying time. These two contributions give, by similar algebra to that of \S\ref{outsidesec},
\begin{align}
    \psi_n(x,t) &\supset \Circled{$\bar{R}$} + \Circled{$\bar{L}$} \notag \\
    &= \mathcal{N}_n \left( \sqrt{\frac{k_n (\bar{z}_0^R)}{k_n(x)}} \times \frac{1}{\sqrt{k_n(\bar{z}_0^R)}} e^{i \bar{F}_+/\hbar} + \sqrt{\frac{k_n (\bar{z}_0^L)}{k_n(x)}} \times \frac{1}{\sqrt{k_n(\bar{z}_0^L)}} e^{i \bar{F}_-/\hbar} \right) \left[ 1 + \mathcal{O}(\hbar) \right] \notag \\
    &= \left( \Circled{$+$} + \Circled{$-$} \right) e^{-i E_n t/\hbar} \notag \\
    &= \tilde{\psi}_n(x) \, e^{-i E_n t/\hbar} \left[ 1 + \mathcal{O}(\hbar) \right] \,, \\
    \Circled{$\pm$} &\equiv \frac{\mathcal{N}_n}{\sqrt{k_n(x)}} \exp \left( \pm \frac{i}{\hbar} \int_{c}^x k_n \mp \frac{i\pi}{4} \right) \left[ 1 + \mathcal{O}(\hbar) \right] \,, \label{plusminuscontri}
\end{align}
where, recall, $\tilde{\psi}_n$ was defined in \eqref{WKBpsin} and here $x$ lies in the classically allowed region. Just as in \S\ref{outsidesec}, after some time $t_1$ (namely $t_{c \to x}$), the right-mover turns into a left-mover, but its contribution to $\psi$ is continuous: $\Circled{$\bar{R}$} = \Circled{$+$} \to \Circled{$\bar{L}$} = \Circled{$+$}$. Similarly, after time $t_2$ ($t_{x \to a}$) the left-mover becomes a right-mover, with $\Circled{$\bar{L}$} = \Circled{$-$} \to \Circled{$\bar{R}$} = \Circled{$-$}$. After $t_3 = t_1 + t_n/2$ the left-mover becomes a right-mover again, and so on. Therefore, the \Circled{$+$} contribution always reaches $x$ with positive momentum, oppositely for the \Circled{$-$} contribution. Hence
\begin{align}
    \psi_n(x,t) \supset \tilde{\psi}_n(x) \, e^{-i E_n t/\hbar} \left[ 1 + \mathcal{O}(\hbar) \right] \quad \text{for all } t > 0 \,. \label{RLexchange}
\end{align}
So these two solutions, which simply oscillate back and forth in the classically allowed region, reproduce the correct stationary state behavior \eqref{psixtint3} when $0 \leq t \ll 1/\Gamma_n$.

\begin{figure}[!ht]
\centering
\includegraphics[width=300pt]{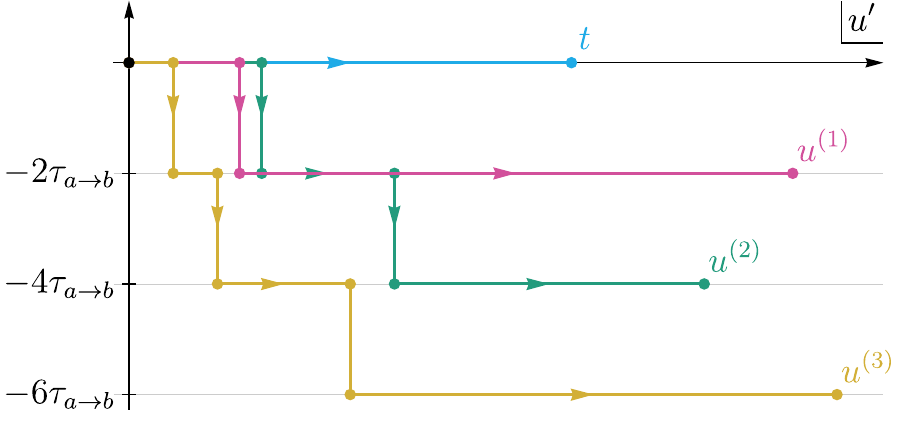}
\caption{Some of the complexified time saddles that contribute to the wave function $\psi_n(x,t)$ inside the false vacuum. On the real line (yellow curve), the solution simply oscillates back and forth in the classically allowed region, reaching $x$ at time $t$. Every time this solution reaches the right turning point $a$ of the false vacuum (see Fig.~\ref{tunneling_setup}), it can go through a negative imaginary time evolution of length $2 \tau_{a \to b}$, after which it is back at $a$ and can continue oscillating. The contribution of each saddle will be calculated by analytic continuation from the appropriate portion of the half-line \eqref{tstarell}. (The appropriate deformations to go around the WKB turning points are implied.)}
\label{complexpaths}
\end{figure}

\vspace*{\baselineskip}
\noindent Now we consider other types of solutions that end on half-lines in the complex $u'$-plane with
\begin{equation} \label{tstarell}
    \Im u^{(\ell)} = -2 i \ell \tau_{a \to b} \,, \quad \ell \in \mathbb{N} \,.
\end{equation}
They are constructed from the oscillating solutions described above by allowing for a negative imaginary time evolution of an integer multiple of $2\tau_{a \to b}$ each time the semiclassical particle reaches the classical turning point $a$. During each such imaginary time segment, the particle travels across the barrier (in the inverted potential) and back: these are finite-time and finite-energy analogs of Sidney's \enquote{bounces} in this formalism. After that either some more real-time evolution ensues, or more bounces. We illustrate a few of such time contours in Fig.~\ref{complexpaths}.

Consider a particular solution that ends at $u^{(\ell)}$ on the half-line \eqref{tstarell}. Its contribution to $\psi_n(x,u^{(\ell)})$ is one of the oscillating contributions \Circled{$\pm$} computed in \eqref{plusminuscontri}, times a suppression factor from the bounces, up to a normalization\footnote{To be sure, $\mathcal{N}_\ell$ is the normalization $\mathcal{N}_P$ we left unspecified in the general formula \eqref{masterformula}. We will show $p=1/2$ further on.} $\mathcal{N}_\ell = p^\ell$ and a factor\footnote{Originating (\S\ref{underbarriersec}) from going clockwise around the turning point $b$ when moving in the negative imaginary time direction. Without the imaginary time transition, the semiclassical prefactor $1/\sqrt{\gamma}$ in \eqref{masterformula} received a factor $e^{-i\pi/2}$ from going around $a$ in the lower half plane. Now instead, it receives a factor $e^{-i \pi/4} \times e^{i \pi/2} \times e^{i \pi/4} = - e^{-i\pi/2}$.} $(-1)^\ell$:
\begin{align} \label{bouncingcontr}
    \psi_n(x,u^{(\ell)}) &\supset (-p \, e^{-2 S_n/\hbar} )^\ell \, \Circled{$\pm$} \, e^{-i E_n u^{(\ell)}/\hbar} \,,
\end{align}
where, recall,
\begin{equation}
    S_n = \int_{a}^{b} \kappa_n
\end{equation}
is the barrier integral at energy $E_n$. The important observation now is that there are \emph{many} such solutions, giving identical contributions to $\psi_n(x,u^{(\ell)})$, when $\Re u^{(\ell)}$ is large. If the semiclassical particle reaches $a$ a total of $N$ times, then at any of them the particle could have made an imaginary time transition, with a total of $\ell$ transitions. There are
\begin{equation} \label{combinatorialfactor}
    C_\ell(N) = \binom{N - 1 + \ell}{\ell}
\end{equation}
such paths. This gives a strong enhancement at large $N$. For the \Circled{$-$} solutions, the number of times $N_-$ that $a$ is reached is related to $u$ by
\begin{equation}
    N_- = \left\lceil \frac{\Re u^{(\ell)} - t_{x \to a}}{t_n} \right\rceil
\end{equation}
while for the \Circled{$+$} solutions it is
\begin{equation}
    N_+ = \left\lceil \frac{\Re u^{(\ell)} - \left( t_n/2 + t_{c \to x} \right)}{t_n} \right\rceil
\end{equation}
so that, all together,
\begin{equation}
    \psi(x,u^{(\ell)}) \supset \left( -p \, e^{-2 S_n/\hbar} \right)^\ell \left( C_\ell(N_-)\, \Circled{$-$} + C_\ell(N_+)\, \Circled{$+$} \right) e^{-i E_n u^{(\ell)}/\hbar} \,.
\end{equation}
Finally, we must analytically continue
$u^{(\ell)} \to t$. At this point Sagredo might be at a loss, because $N_\pm$ are not analytic functions of $u^{(\ell)}$. Salviati might reason, though, that for large $\Re u^{(\ell)}/t_n \gg \ell$, the difference of at most one between $N_+$ and $N_-$ is unimportant, and we can approximate both combinatorial factors as
\begin{equation}
    C_\ell(N_+) \approx C_\ell(N_-) = \frac{\left( \Re u^{(\ell)}/t_n \right)^\ell}{\ell!} \left[ 1 + \mathcal{O} \left( \ell^2/N \right) \right] \,, \label{CNplusCNminus}
\end{equation}
which to his frustration, Salviati realizes is still not analytic. With renewed hope Sagredo proceeds nevertheless, summing over all solutions and finding
\begin{align}
    \psi_n(x,u^{(\ell)}) &\approx \tilde{\psi}_n(x) \, e^{-i E_n u^{(\ell)}/\hbar} \sum_{\ell=0}^\infty \frac{1}{\ell!} \left(-p \frac{\Re u^{(\ell)}}{t_n} e^{-2 S_n/\hbar} \right)^\ell [1+\mathcal{O}(\hbar)] \label{tuscansum} \\
    &= \tilde{\psi}_n(x) \, e^{-i E_n u^{(\ell)}/\hbar} \exp \left( -p \Gamma_n \Re u^{(\ell)} \right) [1+\mathcal{O}(\hbar)] \,. \label{Sagredoalmost}
\end{align}
He then observes that replacing $\Re u^{(\ell)}$ with $u^{(\ell)}$ would introduce an exponentially small correction, as $\Gamma_n$, defined in \eqref{Gammandef}, is exponentially small. \eqref{Sagredoalmost}, with its undetermined $\mathcal{O}(\hbar)$ correction, is too crude to capture such an effect anyway, so that the replacement is justified. After this, the twosome sends $u^{(\ell)} \to t$ and finds
\begin{equation} \label{psixtinsidePIp}
    \psi_n(x,t) \approx \tilde{\psi}_n(x) \, e^{-i E_n t/\hbar} \times e^{-p \Gamma_n t} [1+\mathcal{O}(\hbar)] \,, \quad \text{for } 0 \leq t \lesssim 1/\Gamma_n \,.
\end{equation}
For $t \lesssim 1/\Gamma_n$, the dominant contributions to the infinite sum in \eqref{tuscansum} come from $\ell = \mathcal{O}(1)$ bounces, and it was indeed consistent to neglect the corrections to \eqref{CNplusCNminus}.

Only $p$ remains to be determined. It can be fixed by conservation of probability, using the known behavior of the wave function in the free region $x > b$. At early times $t_n \lesssim t \ll 1/\Gamma_n$, which has an overlap with the time frame in \eqref{psixtinsidePIp}, we computed the wave function in the free region via a \emph{normalized} semiclassical saddle to be
\begin{equation} \label{psiXToutsideEARLY}
    \psi_n(x,t) = \mathcal{N}_n \frac{(-1)^n}{\sqrt{k_n(x)}} \exp \left( \frac{i}{\hbar} \int_{b}^{x} k_n + \frac{i \pi}{4} \right) e^{-S_n/\hbar} e^{-i E_n t/\hbar} \left[ 1 + \mathcal{O}(\hbar) \right] \,, \quad \text{for } t_n \lesssim t \ll 1/\Gamma_n \,.
\end{equation}
On the one hand, we know from \eqref{psixtinsidePIp} that the probability to be inside the false vacuum at time $t$ is
\begin{align} \label{PFVt}
    P_\textsf{FV}(t) &= \int_{-\infty}^{x} \di x' \, \abs{\psi_n(x',t)}^2 \approx e^{-2p \Gamma_n t} \,,
\end{align}
where we used that the main contribution to \eqref{PFVt} comes from inside the false vacuum where \eqref{psixtinsidePIp} is valid. Therefore, for $t_n \lesssim t \ll 1/\Gamma_n$,
\begin{equation}
    -\dot{P}_\textsf{FV}(t) \approx 2p \Gamma_n \,.
\end{equation}
On the other hand, from the Schr\"odinger equation it follows that
\begin{equation} \label{latetimecurrent}
    -\dot{P}_\textsf{FV}(t) = -\frac{i \hbar}{2m} \left( \psi_n^* \partial_x \psi_n - \psi_n \partial_x \psi_n^* \right)\Big|_{x,t} \,.
\end{equation}
Using \eqref{psiXToutsideEARLY} and recalling \eqref{Nnnorm}, we find
\begin{equation} \label{Gammanoutside}
    -\dot{P}_\textsf{FV}(t) \approx \Gamma_n \quad\implies\quad p = \frac{1}{2} \,.
\end{equation}
With this, the semiclassical path integral computation \eqref{psixtinsidePIp} matches the WKB result \eqref{psixtint3}.

\paragraph{A word about dominant vs.\ subdominant contributions}
Bouncing solutions give contributions \eqref{bouncingcontr} to $\psi$ that are exponentially suppressed as $\hbar \to 0$ compared to the two non-bouncing contributions \eqref{RLexchange}. More specifically, individual bouncing contributions are subdominant compared to the uncomputed $[1 + \mathcal{O}(\hbar)]$ correction to \eqref{RLexchange}, so that it would appear meaningless to keep them. We noted, however, that at late times the bouncing contributions receive a combinatorial enhancement (because there are many giving the same contribution), so that their relative magnitude compared to the non-bouncing contribution is
\begin{equation}
    \frac{1}{\ell!} \left( \frac{t}{2 t_n} e^{-2 S_n/\hbar} \right)^\ell = \frac{1}{\ell!} \left( \frac{\Gamma_n t}{2} \right)^\ell \,.
\end{equation}
So for times of order $t = \mathcal{O}(1/\Gamma_n)$, the collective contribution of all bouncing solutions with $\ell = \mathcal{O}(1)$ competes with the non-bouncing contribution and we must keep all.

\paragraph{A word about zero modes and negative modes}
In the standard \cite{Coleman:1978ae} discussion, the imaginary time version $t' \to -i\tau$ of the fluctuation operator \eqref{fluctuationoperator} around the (real) bounce solution $z_b(\tau)$, call it $\mathcal{F}_\tau$, has a zero mode. It is proportional to $\partial_\tau z_b$: this is a solution to $\mathcal{F}_\tau \left( \partial_\tau z_b \right) = 0$, approximately satisfying the Dirichlet boundary conditions $\partial_\tau z_b(\pm T/2) = 0$ with $T$ the total imaginary time, which is taken to infinity. Instead of integrating over the perturbative zero mode, one introduces a collective coordinate that corresponds to the center of the bounce \textdash which could have been anywhere in $(-T/2,T/2)$ \textdash and integrates over this, which gives an enhancement factor proportional to $T$. One further concludes that $\mathcal{F}_\tau$ has one negative mode, because the zero mode $\partial_\tau z_b$ has one node. This mode is rotated by 90 degrees to obtain a convergent Gaussian integral, yielding a crucial factor of $i$ to conclude that the bounce describes tunneling. One also argues that effectively only half of the would-be negative mode should be integrated over, giving a factor $1/2$ (see also \cite{Andreassen:2016cvx}).

In the present formalism one can simply evaluate
\begin{equation}
    \det \mathcal{F} \propto \gamma(t) = \frac{\dot{\bar{z}}(t)}{\dot{\bar{z}}(0)} \,,
\end{equation}
which is non-zero, so $\mathcal{F}$ does not have a zero mode. This is because (the components of) our initial state \eqref{decomppsit0} have a finite energy above the bottom of the false vacuum. More generally, $\bar{z}$ is complex in our case and $\mathcal{F}$ has complex eigenvalues (and is not Hermitian). We were not concerned with this because the master formula \eqref{masterformula} we obtained is well-defined for complex backgrounds and yields a solution to the Schr\"odinger equation to $\mathcal{O}(\hbar^1)$. Although there is no zero mode in our calculation, we still obtain an enhancement factor proportional to the total time (which is large but finite), coming from a large amount of inequivalent bounces giving identical contributions; morally this is very similar to integrating over the location of the center of the bounce. Regarding the negative mode, strictly speaking it is also absent in our case. But one could say that it entered implicitly when we instructed our contours in the complex $u'$-plane to avoid the WKB turning points. For a solution with $\ell$ finite-time bounces, this introduced a crucial factor $(-1)^\ell$, causing the wave function inside the false vacuum to \emph{decay} exponentially at late times. Finally, in our case, the factor $(1/2)^\ell$ for an $\ell$-bounce solution is a normalization of the path integral that in this formalism accompanies every (initially) subleading contribution. We determined it by demanding a local conservation of probability, equating the known, normalized expression for the wave function outside the false vacuum at intermediate times $t_n \lesssim t \ll 1/\Gamma_n$ to the unnormalized wave function inside the false vacuum in the same time frame.

\subsubsection{Back to the outside, at late times} \label{backoutsidesec}
Returning to the wave function outside the metastable region, it is now a simple matter to obtain the exponential damping factor in the WKB result \eqref{psiXToutside} from the path integral: one merely sums over all solutions which have the possibility to \enquote{bounce} in the imaginary time direction every time they reach $a$, following the procedure of \S\ref{insidesec}. At the final encounter with $a$, the particle goes through the barrier (once), arriving at $x > b$ at time $u^{(\ell)}$, which is continued to $t$. Apart from the exponential suppression $(e^{-2 S_n/\hbar})^\ell$, each $\ell$-bounce solution receives a factor $(-1)^\ell \times (1/2)^\ell$ compared to the $0$-bounce solution, and their multiplicity is effectively $(t/t_n)^\ell/\ell!$. This gives \eqref{psiXToutside}.

\subsubsection{Comparison of various methods in an explicit example} \label{numericssec}
\begin{figure}[ht!]
    \centering
    \includegraphics[width=330pt]{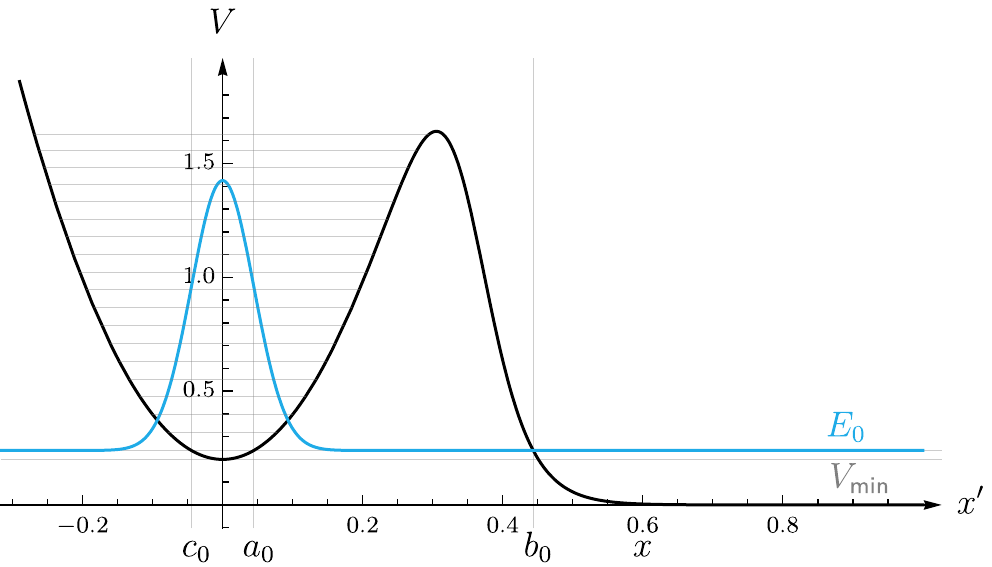}
    \caption{Tunneling potential \eqref{tunnelingform} with parameters $m=1$, $\omega=2\pi$, $x_T=44/125$, $\Delta x=33/500$ and $V_0 = 1/5$. The initial state (blue) is the Gaussian \eqref{HOintialstate} \textdash a good approximation to the resonant ground state of the false vacuum \textdash shown for $\hbar = 1/80$. For this value there are 19 resonant states.}
\label{fig:tunnelingpotential}
\end{figure}

\noindent In this section we study an explicit tunneling scenario, with potential
\begin{equation} \label{tunnelingform}
    V(y) = \left(V_0+\frac{1}{2} m \omega ^2 y^2\right) \times \frac{1}{2} \left[ 1 - \tanh \left(\frac{y-x_T}{\Delta x} \right) \right] \,, \qquad \text{(Fig.~\ref{fig:tunnelingpotential})}
\end{equation}
and compare four methods of computing the wave function $\psi(x,t)$ for $x > b_0$ outside the false vacuum. For the initial state, we choose a Gaussian centered at the minimum $x_\textsf{min}$ of the false vacuum with a width corresponding to the ground state of the quadratic truncation of the potential around the minimum:
\begin{equation} \label{HOintialstate}
    \psi(x_0,0) = \mathcal{N}_0 \exp \left( - \frac{(x_0-x_\textsf{min})^2}{2d^2} \right) \,, \quad \mathcal{N}_0^2 = \frac{1}{d \sqrt{\pi}} \,, \quad d^2 = \frac{\hbar}{m \Omega} \,, \quad \Omega = \sqrt{V''(x_\textsf{min})/m} \,.
\end{equation}
(Notice $x_\textsf{min} \approx 0$, $\Omega \approx \omega$.) As we send $\hbar \to 0$ this becomes a good approximation to the resonant ground state of the metastable region. In this case the relevant classical oscillation time is $2\pi/\omega$, and we will limit ourselves to \enquote{early} times
\begin{equation} \label{timeframeofinterest}
    0 \leq t < \mathcal{O}(1/\omega) \,,
\end{equation}
because the exponentially late time $t \sim 1/\Gamma_0$ at which the wave function starts to decay considerably is difficult to probe numerically. Recall that in \S\ref{outsidesec} we studied the time evolution of \emph{excited} resonant states, so we cannot copy that section verbatim (although the analytic results are identical up to $\mathcal{O}(1)$ factors).

The four methods are
\begin{enumerate}
    \item WKB.
    \item Indirect (complex time) semiclassical path integral.
    \item Numerical solution to the Schr\"odinger equation.
    \item Direct (real time) semiclassical path integral.
\end{enumerate}
The first two methods are analytic and proceed along the lines of \S\ref{outsidesec}. In particular, we will argue they match. As in our computation in \S\ref{outsidesec}, however, 1.\ and 2.\ are only expected to become valid approximations after a time of order $1/\omega$. After this time, components of the Gaussian initial state with energy above the barrier will have decayed, and we are left only with the resonant excited state components. It turns out that in the setup above, as $\hbar \to 0$, the excited under-barrier components dominate the signal $\psi(x,t)$ for a time $1/\omega \lesssim t \lesssim 1/\Gamma_* \sim (1/\omega) e^{2 S(E_*)/\hbar}$ for a certain critical energy $E_*$ below the top of the barrier. We can understand this as follows: for $t \gtrsim 1/\omega$, we have, to good approximation,
\begin{equation} \label{psixtresonantdecomp}
    \psi(x,t) = \sum_{n=0}^{N_\textsf{max}} \langle \psi_n | \psi_{t=0} \rangle \, \psi_n(x,t) \,,
\end{equation}
with $\psi_n$ the time evolution of the resonant states outside the well, \eqref{psiXToutside}. For $n = 0$, the coefficient $\langle \psi_n | \psi_{t=0} \rangle$ is one up to an exponentially small correction, coming from the fact that $\psi_{t=0}$ stops being a good approximation to the resonant ground state at some location $x_*$ inside the barrier (in our case, it happens when the higher-order corrections to the quadratic truncation of the potential become important). For the same reason, coefficients for $n \geq 1$ are not zero but rather\footnote{To estimate the overlap, we used that a significant contribution to $\langle \psi_n | \psi_{t=0} \rangle$ comes from the region $x < x_*$. In this region, $H \psi_n(x) = E_n \psi_n(x)$ and $H \psi_{t=0}(x) = E_0 \psi_{t=0}(x)$ to good approximation, and then we expressed $\langle H \psi_n | \psi_{t=0} \rangle_{x < x_*}$ in terms of $\langle \psi_n | H \psi_{t=0} \rangle_{x < x_*}$ plus boundary terms.}
\begin{align} \label{overlapsemiclassical}
    \langle \psi_n | \psi_{t=0} \rangle \approx \frac{\hbar}{2m} \frac{\kappa_n(x_*) - \kappa_0(x_*)}{E_n - E_0} \psi_n(x_*) \psi_{t=0}(x_*) \sim \exp \left( - \frac{1}{\hbar} \int_{a_n}^{x_*} \kappa_n - \frac{1}{\hbar} \int_{a_0}^{x_*} \kappa_0 \right) \,,
\end{align}
where we will be concerned with the leading semiclassical behavior only. (Here we have assumed $a_n < x_*$; otherwise the first term would be absent.) The wave functions $\psi_n(x,t)$ themselves involve a factor $e^{-S_n/\hbar}$. Putting everything together, we find the following condition for the $n$th term in \eqref{psixtresonantdecomp} to dominate over the $n=0$ contribution in the semiclassical limit:
\begin{equation} \label{dominantcondition}
    2 \int_{a_n}^{x_*} \kappa_n + \int_{x_*}^{b_n} \kappa_n < \int_{x_*}^{b_0} \kappa_0 \,.
\end{equation}
(When $a_n > x_*$, the factor 2 becomes 1.) For low-lying excited states, the inequality is not satisfied and so these states are irrelevant in the sum \eqref{psixtresonantdecomp} at all times. There is a first level $n_*$ corresponding to energy $E_*$ for which the inequality starts becoming valid, however. Therefore, we must wait a time $1/\Gamma_*$ for this component to decay, before remaining with the ground-state component only. For the potential \eqref{tunnelingform}, taking $x_* = 0.23$, we find $E_* = 0.7$.\footnote{We defined $x_*$ by the condition $H \psi_{t=0}/E_0 \psi_{t=0}\big|_{x_*} - 1 = \varepsilon_t = 1/10$, and verified that changing the threshold $\varepsilon_t$ does not affect the results significantly.} It is clear that in general, one shortens the time $1/\Gamma_*$ by having the initial state track the resonant ground state deeper inside the barrier.

The earlier-time behavior $t \lesssim 1/\omega$ is dominated by components with energy above or close to the top of the barrier, and is not captured by the above discussion. Nevertheless, method 4., introduced in \S\ref{mainsec}, applies to this regime as well and can be compared to the numerical solution. At these early times we will observe exchanges of dominance between saddle points, illustrating the general ideas of \S\ref{mainsec}. In the end, all methods agree in the regimes where they should.

\paragraph{WKB}
(See also \cite{garg2025,JKNtoappear}.) For low-lying states, the WKB approximation with linear turning points is inapplicable in the classically allowed region. Instead, one starts by approximating the potential as purely quadratic at the bottom of the false vacuum, solving the Schr\"odinger equation for \emph{continuous} values of the energy and demanding that the wave function decays to one side. This approximation can be trusted deep inside the barrier, where, for general values of the energy, the wave function has a growing branch. By matching the asymptotic expansion to a standard WKB form under the barrier, the solution can be extended to the free region. The final result for the time evolution of the $n$th resonant state outside the well is
\begin{align}
\label{eq:psi_n_WKB}
    & \psi_n(x,t) = \sqrt{\frac{m\omega}{2\pi g_n}} \frac{(-1)^n}{\sqrt{k_n(x)}} \exp \left( \frac{i}{\hbar} \int_{b_n}^x k_n + \frac{i \pi}{4} \right) e^{-S_n/\hbar} e^{-i E_n t/\hbar} e^{-\Gamma_n t/2} \left[ 1 + \mathcal{O}(\hbar) \right] \,, \\
\label{eq:gamma_n_WKB}
    & \Gamma_n = \frac{\omega}{2\pi g_n } e^{-2 S(E_n)/\hbar} \,, \quad g_n = \frac{1}{\sqrt{2\pi}} \left( \frac{e}{n+1/2} \right)^{n+1/2} n! \,, \quad E_n = V_\textsf{min} + \hbar \omega \left( n + 1/2 \right) \,.
\end{align}
So the only difference with the result for highly excited states \eqref{psiXToutside}, is that $t_n \to g_n t_n = 2 \pi g_n/\omega$, where $g_n \to 1$ as $n \to \infty$, thereby matching onto \eqref{psiXToutside} for $n \gg 1$ where it is valid. For the ground state, $g_0 = \sqrt{e/\pi} \approx 0.93$. Beware that here, the barrier integrals $S_n$ are evaluated between the inside and outside WKB turning points $a_n$ and $b_n$ satisfying $V(a_n) = V(b_n) = E_n$, even for the ground state.

\paragraph{Indirect (complex time) method} We will confine ourselves to the path integral computation of the time evolution of the resonant ground state, i.e.\ level $n = 0$. Once again, we will evaluate the wave function $\psi_0(x,u)$ at an external position $x > b$ by constructing a suitable complexified time contour. In this paragraph we will refer to $b$ as the outer classical turning point satisfying $V(b) = V_\textsf{min}$, and we set $x_\textsf{min} = 0$, $\Omega = \omega$ to avoid clutter. This differs slightly from the $b_0$ introduced above which had $V(b_0) = E_0 = V_\textsf{min} + \hbar \omega/2$, but it is important to keep the difference in mind to match the WKB result including $\mathcal{O}(1)$ factors.

We will focus only on the leading saddle at times earlier than $1/\Gamma_0 \sim (1/\omega) e^{2 S_0/\hbar}$, as the multi-bounce contributions can be accounted for as in \S\ref{insidesec}. Evaluating the saddle points at a complex final time $u= t - i\tau$, where $t, \tau \gg 1/\omega$, the leading trajectory $\bar{z}(u')$ proceeds along three segments:
\begin{enumerate}
\item Real-time rotation ($0 \to u_c$): starting at a complex position near but not at the origin $\bar{z}(0) = \bar{z}_0$, the trajectory revolves in the quadratic well, $\bar{z}(t') = \bar{z}_0 \, e^{i\omega t'}$, until reaching a strictly real point $\bar{z}(u_c) = \abs{\bar{z}_0}$ on the side of the barrier. The boundary condition $\dot{\bar{z}}(0) = i\omega \bar{z}_0$ imposes an effective classical energy $E = V_\textsf{min}$ (not $V_\textsf{min} + \hbar \omega/2$).
\item Imaginary transmission ($u_c \to u_c - i\tau$): the trajectory travels in negative imaginary time under the barrier, remaining real-valued, and emerging at the zero-energy turning point $\bar{z}(u_c - i\tau) = b$. Notice that $\tau$ is finite for $\abs{\bar{z}_0} \neq 0$.
\item Real-time escape ($u_c - i\tau \to u$): the particle rolls away from the barrier in real time to reach $\bar{z}(u) = x$.
\end{enumerate}

Evaluating the classical action across the three segments yields
\begin{equation}
\label{eq:action_segments}
S_1 = i\frac{m\omega}{2}(\abs{\bar{z}_0}^2 - \bar{z}_0^2) \,, \quad S_2 = i\int_0^{b} \kappa_\mathsf{min} - i\frac{m\omega}{2}\abs{\bar{z}_0}^2 \equiv iS_\mathsf{min} - i\frac{m\omega}{2}\abs{\bar{z}_0}^2 \,, \quad S_3 = \int_{b}^x k_\mathsf{min} \,,
\end{equation}
where $\kappa_\mathsf{min}(x) = \sqrt{2m(V(x)-V_\textsf{min})}$ and $k_\mathsf{min}(x) = \sqrt{2m(V_\textsf{min}-V(x))}$. Summing these components and combining them with the initial state exponent cancels the initial position $\bar{z}_0$ entirely:
\begin{equation}
\label{eq:classical_exponent}
\exp\left( \frac{i}{\hbar}S[\bar{z}] - \frac{f(\bar{z}_0)}{\hbar} \right) = \exp\left( -\frac{S_\mathsf{min}}{\hbar} + \frac{i}{\hbar}\int_{b}^x k_\mathsf{min} \right) \,.
\end{equation}

To evaluate the one-loop prefactor, $\gamma(u)^{-1/2}$, we note that $\gamma(u') = \dot{\bar{z}}(u')/\dot{\bar{z}}(0)$ satisfies the normalized fluctuation equation. The final velocity is given by $k_\mathsf{min}(x)/m$, while the initial velocity can be expressed as
\begin{equation}
\label{eq:initial_vel}
\dot{\bar{z}}(0) = i\omega \bar{z}_0 = i\omega \abs{\bar{z}_0} e^{-i\omega u_c} = i\omega \abs{\bar{z}_0} e^{\omega \tau} e^{-i\omega u} e^{i\omega t_{b \to x}}.
\end{equation}
Notice that there is no zero mode, because $\bar{z}_0 \neq 0$. To calculate $\abs{\bar{z}_0} e^{\omega\tau}$ at large $\tau$, we regularize the under-barrier travel time $\tau$ near the origin (analogously to \cite{garg, garg2025, Marino:2021lne}):
\begin{equation}
\label{eq:integral_A}
A \equiv \int_0^{b} \di x \left( \frac{m\omega}{\sqrt{2m(V(x)-V_\textsf{min})}} - \frac{1}{x} \right) .
\end{equation}
This gives $\omega \tau \approx A + \log(b/\abs{\bar{z}_0})$, so that $\abs{\bar{z}_0} e^{\omega \tau} = b \, e^A$. Combining the classical and one-loop results and continuing $u\to t$, we obtain the wave function
\begin{equation}
\label{eq:ground_state_path_int}
\psi_0(x,t)= \mathcal{N}_0 \sqrt{ \frac{m\omega b}{k_\mathsf{min}(x)} }\exp\left( -\frac{S_\mathsf{min}}{\hbar} + \frac{i}{\hbar}\int_{b}^x k_\mathsf{min} -i\frac{\omega t}{2} + i\frac{\omega t_{b \to x}}{2} + \frac{A}{2} + \frac{i \pi}{4}\right) \left[ 1 + \mathcal{O}(\hbar) \right].
\end{equation}

As we have stressed, standard WKB calculations evaluate phase integrals at the zero-point energy $E_0 = V_\textsf{min} + \hbar\omega/2$ rather than $E = V_\textsf{min}$. Expanding $k_0(x') \approx k_\mathsf{min}(x') + (\hbar m \omega/2)/k_\mathsf{min}(x')$, the integral shifts as
\begin{equation}
\label{eq:phase_shift}
\frac{1}{\hbar}\int_{b_0}^x k_0 \approx \frac{1}{\hbar}\int_{b}^x k_\mathsf{min} + \frac{\omega t_{b \to x}}{2} \,.
\end{equation}
This shift absorbs the $t_{b \to x}$ phase generated by the one-loop factor. Next, we relate the zero-energy bounce action $S_\mathsf{min}$ to the finite-energy action $S_0$. Evaluating the difference between $\kappa_\mathsf{min}$ and $\kappa_0$ in the quadratic region yields
\begin{equation}
\label{eq:action_shift}
S_\mathsf{min} = S_0 + \frac{\hbar}{2} \left( A + \log (2b/d) + \frac{1}{2} \right) \,.
\end{equation}
Upon substituting \eqref{eq:action_shift} into \eqref{eq:ground_state_path_int}, the parameters $A,b$ cancel out and the resulting wave function reproduces the WKB result \eqref{eq:psi_n_WKB} for the ground state $n=0$.

\paragraph{Numerical solution}
To numerically integrate the Schr\"odinger equation, we represent the real positions in a discretized interval $[x_L, x_R]$. In this way the spectrum becomes discrete, with a UV cutoff set by the minimum step $\delta x$.
We introduce a uniform grid
\begin{equation}
x_j = x_L + j \,\delta x \, ,
\qquad
\delta x = \frac{x_R-x_L}{N-1} \, ,
\end{equation}
with $j \in\{0,\dots,N-1\}$.
The wave function is represented by the vector $(\psi_i(t)) \equiv (\psi(x_i,t))\in\mathbb{C}^N$, while the Hamiltonian becomes a finite-dimensional matrix.
The potential term becomes a diagonal matrix,
\begin{equation}
U_{ij} = V(x_i)\,\delta_{ij} \, ,
\end{equation}
while the kinetic operator is approximated by a second-order finite-difference Laplacian, represented as a tridiagonal matrix
\begin{equation}
T_{ij} = -\frac{\hbar^2}{2m}\,
\frac{1}{\delta x^2}
\left(
\delta_{i+1,j} - 2\delta_{ij} + \delta_{i-1,j}
\right) .
\end{equation}
The full Hamiltonian is therefore
\begin{equation}
H = T + U \, .
\end{equation}
The initial state is discretized and normalized according to
\begin{equation}
\sum_i \abs{\psi_i(0)}^2 \, \delta x = 1 \, .
\end{equation}
Time evolution is implemented through the unitary operator generated by the discretized Hamiltonian,
\begin{equation}
\psi(t+\delta t)
= e^{- i H \delta t/\hbar}\, \psi(t) \, ,
\end{equation}
which we apply iteratively to obtain the full time evolution.

We impose boundary conditions such that the wave function remains constant at $x_L$ and $x_R$. The position of these boundaries must be chosen sufficiently far from the dynamically relevant region. In particular, the (arbitrary) observation point $x=0.6$, which we focus on, in the potential of Fig.~\ref{fig:tunnelingpotential} lies well inside the interval, so that boundary effects are negligible over the time range of interest. At sufficiently late times, however, back-reflection from the boundaries inevitably appears due to the finite size of the box. (An alternative would have been to enforce outgoing boundary conditions to the right by adding a complex absorbing term to the potential; see e.g.\ \cite{RissMeyer1993,JKNtoappear}.)

The robustness of the numerical procedure has been checked by verifying convergence under grid refinement and stability with respect to variations of $x_L$ and $x_R$. Moreover, for the parameter range explored, we compared our discretized evolution with the direct integration of the Schr\"odinger equation using \texttt{NDSolve} in \emph{Mathematica}, finding excellent agreement. The corresponding evolution of $\hbar \log \abs{\psi(x,t)}$ is shown in Figs.~\ref{psi_instantons_80} and \ref{psi_instantons_20} for different values of $\hbar$.

\begin{figure}[h!]
\centering
\includegraphics[width=330pt]{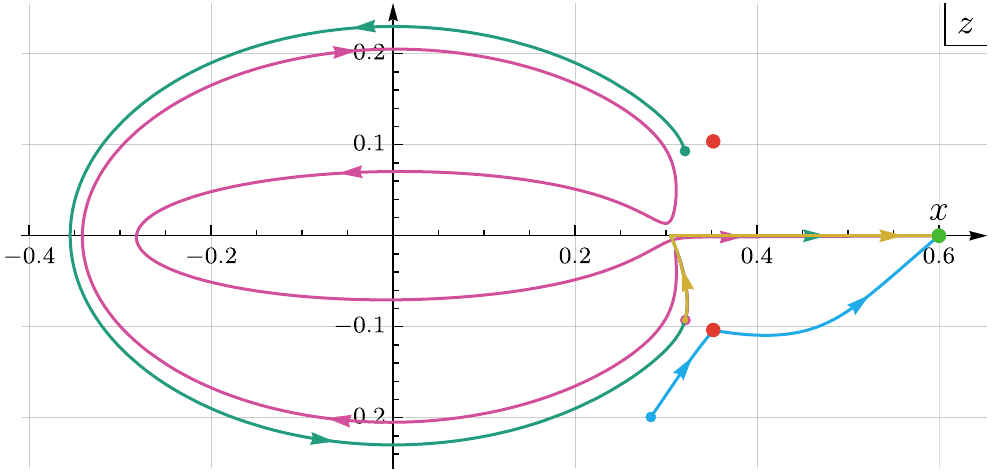}
\caption{Four dominant trajectories $\bar{z}(t')$ determining the leading semiclassical behavior of $\psi(x = 0.6,t)$ at different times $t$. The blue trajectory corresponds to the leading saddle as $t \to 0$ and is shown here shortly before it ceases to exist, around $t = 0.40$. The yellow, green and pink ones are at $t = 1.03$, $t = 1.87$ and $t = 2.64$, respectively. After some complex motion, they are seen to emerge on the real line near the top of the barrier at $x' \approx 0.3$ and slowly roll down to $x$ from there. The two red dots mark poles of the potential.}
\label{instanton_trajectories}
\end{figure}

\begin{figure}[h!]
\centering
\includegraphics[width=350pt]{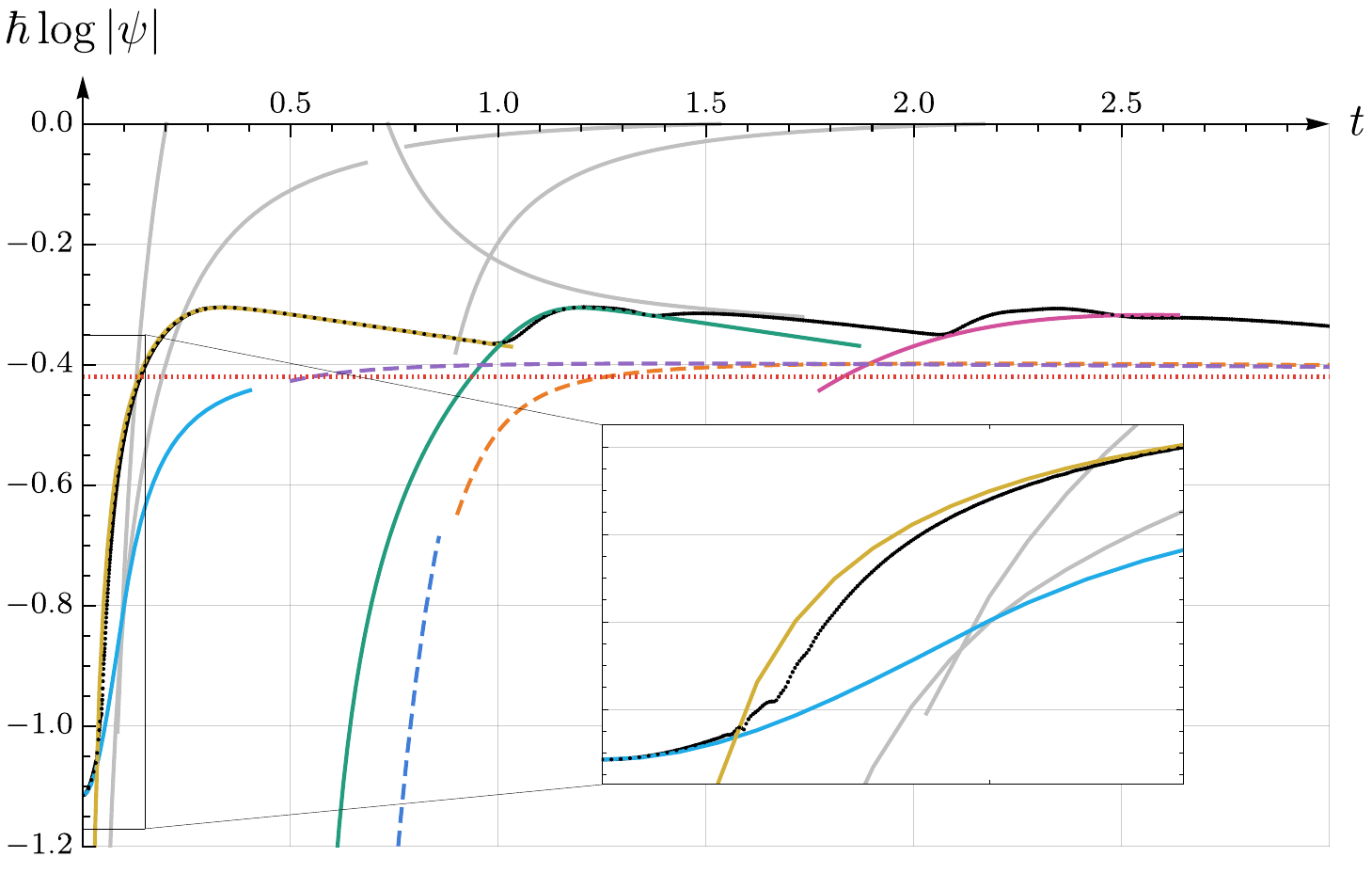}
\caption{Comparison between the real part of the rescaled numerical wave function exponent (black) and individual saddle point contributions at $x = 0.6$ for $\hbar = 1/80$, as a function of time. Colored lines denote the contributing solutions, matching the trajectories in Fig.~\ref{instanton_trajectories}. Gray lines represent non-contributing solutions, while dashed lines indicate sectors that are either subdominant or non-contributing. The red dotted line is the WKB or indirect complex-time result \eqref{eq:psi_n_WKB} for $n = 0$. We explain the discrepancy between this line and the numerical curve in the main text (see also Fig.~\ref{amazing_bumps}). In the bottom-right corner we zoom in on the earliest exchange of dominance of saddles.}
\label{psi_instantons_80}
\end{figure}

\begin{figure}[h!]
\centering
\includegraphics[width=350pt]{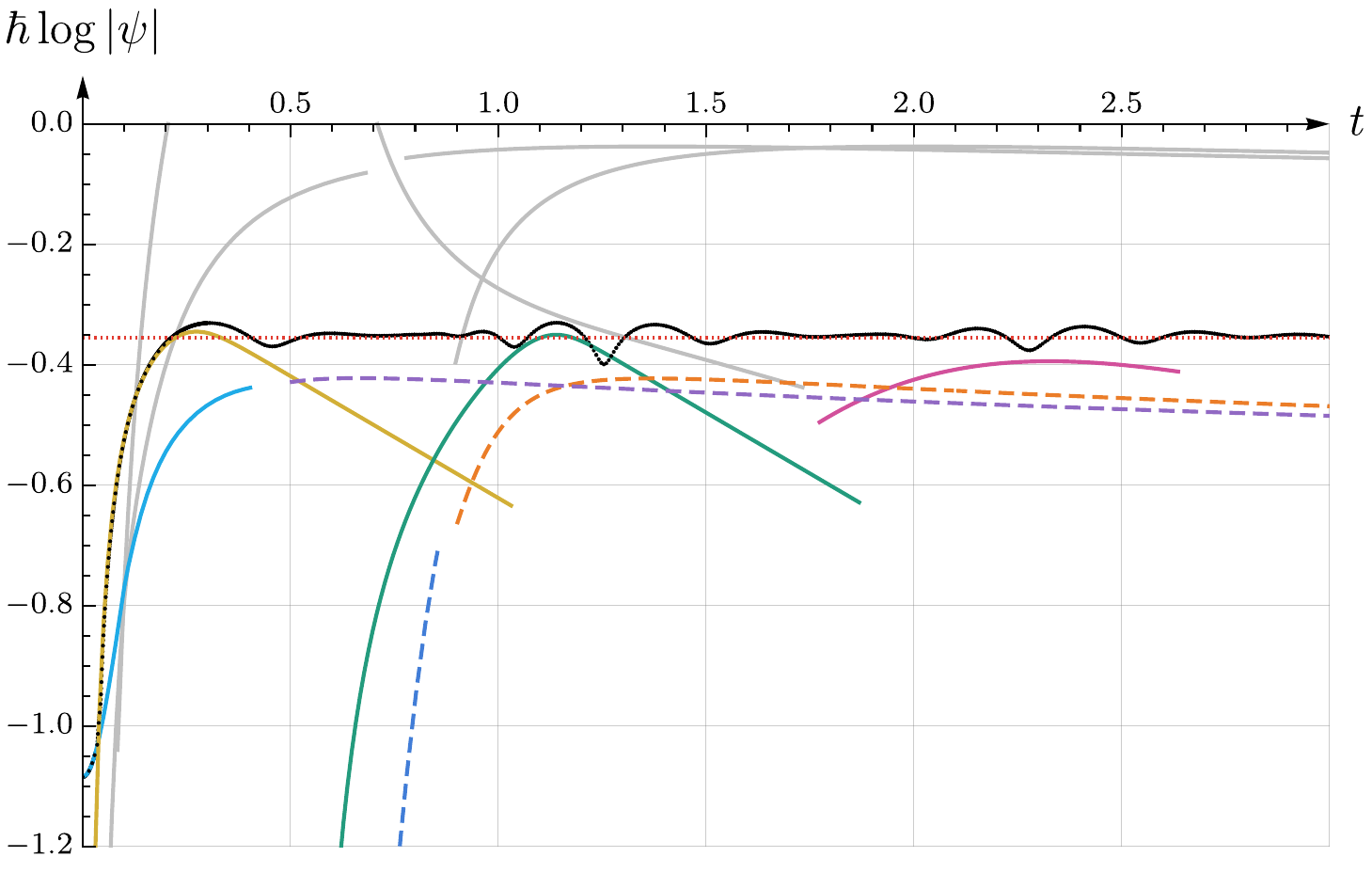}
\caption{Same as Fig.~\ref{psi_instantons_80} but for $\hbar = 1/20$. For this larger value, the agreement with the individual complex saddle point contributions is visibly reduced. It would be instructive to compute a few higher-loop corrections to \eqref{masterformula} and check whether the agreement is restored, as a proof of principle. The value for the wave function obtained from \eqref{eq:psi_n_WKB} for $n = 0$ alone shows a high level of agreement with the numerical solution, in contrast to Fig.~\ref{psi_instantons_80} where $\hbar$ was smaller. For $\hbar = 1/20$, the semiclassical estimate in \eqref{overlapsemiclassical} is not accurate enough and it turns out the $n \neq 0$ contributions in \eqref{psixtresonantdecomp} are all subdominant compared to the $n = 0$ contribution.}
\label{psi_instantons_20}
\end{figure}

\paragraph{Direct (real time) method}  We consider the problem of determining complex saddle points of the path integral that solve the differential equation \eqref{classicalEOM}, subject to the boundary conditions \eqref{initialv} and $\bar{z}(t)=x$, with $f(x)=(x-x_\textsf{min})^2/2d^2$. Once a saddle $(\bar{z}_0,\bar{z})$ is obtained, its contribution to the wave function can be evaluated by substituting it into \eqref{masterformula}. We set $\mathcal{N}_P = 1$ for all saddles. At early times, the dominant contribution comes from the solution that is continuously connected to the unique saddle $\bar{z}=x$ for $t=0$. As time increases, additional solutions to the differential problem emerge. Figure~\ref{instanton_trajectories} shows some dominant trajectories at different times.

The contributions of the various saddles to the path integral are shown in Figs.~\ref{psi_instantons_80} and~\ref{psi_instantons_20}, together with the numerical solution of the Schr\"odinger equation. As expected, the agreement with the numerical solution improves as $\hbar \to 0$. Note that a large number of solutions are either subdominant or do not contribute to the path integral at all.

There are two mechanisms by which we fail to track saddles to earlier or later times. Either a saddle may hit a pole and cease to exist, or its initial condition may become fine-tuned and our numerics fail. The former happens around $t = 0.40$ for the initially dominant solution (blue) and around $t = 1.77$ for the fourth one (pink). The latter happens for the yellow, green and pink curves at the times they are shown in Fig.~\ref{instanton_trajectories} \textdash their (complex) energies approach the local maximum of the potential and the solutions spend a lot of time there. Presumably, these solutions continue to exist for arbitrarily long times.

\begin{figure}[h!]
\centering
\includegraphics[width=350pt]{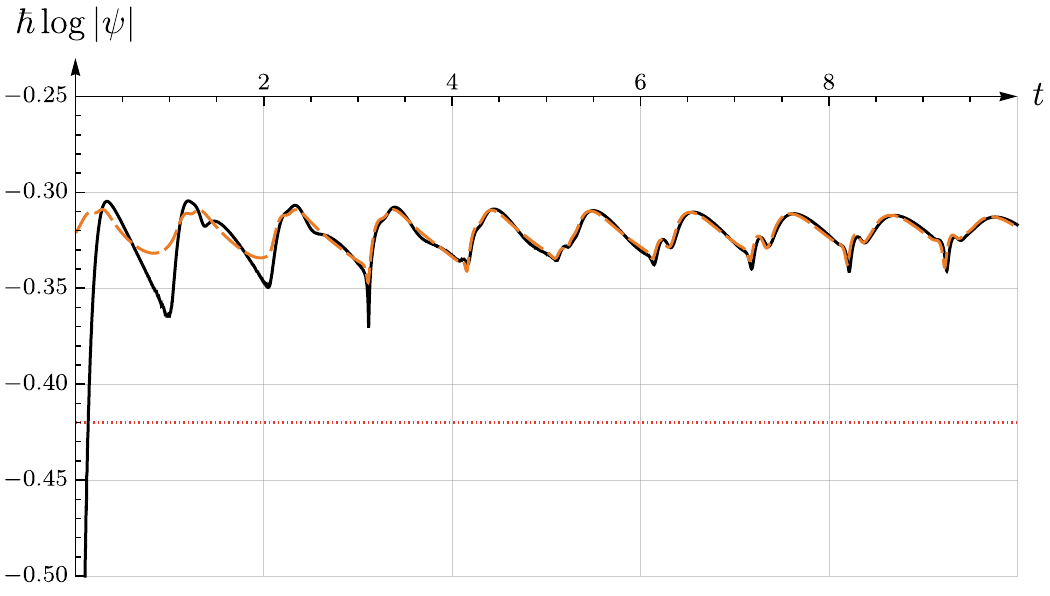}
\caption{Returning to $\hbar = 1/80$, we compare the numerical result (black) to the full sum \eqref{psixtresonantdecomp} (orange dashed), including all the under-barrier excited resonant states. The coefficients $\langle \psi_n | \psi_{t=0} \rangle$ were calculated numerically. After a time $t = \mathcal{O}(1/\omega)$ \textdash when the components of $\psi_{t=0}$ with energy above the barrier have decayed \textdash the agreement is excellent. The ground-state component (red dotted) will become an accurate approximation only after a time of order $1/\Gamma_* = (1/\omega)e^{2 S(E_*)/\hbar}$ with $S(E_*) \sim S(0.7) \approx 0.23$. At this time the last component in \eqref{psixtresonantdecomp} that is dominant compared to the $n = 0$ contribution will have decayed. For $\hbar = 1/80$, $\omega = 1$, we find $1/\Gamma_* \sim 10^{16}$, which is much smaller still than the decay time of the ground state itself, $1/\Gamma_0 \sim 10^{29}$.}
\label{amazing_bumps}
\end{figure}

In Fig.~\ref{psi_instantons_80}, which shows the results for the smallest value of $\hbar$ we considered, we observe bumps with a period of order $1/\omega$ and a typical magnitude of about $-0.3$. These bumps are caused by the admixture of the Gaussian initial state with resonant energy eigenstates with energy $E_* \gtrsim 0.7$, as explained around \eqref{dominantcondition}. We demonstrate this in Fig.~\ref{amazing_bumps} by plotting the full sum \eqref{psixtresonantdecomp}, which matches the numerical evolution for $t \gtrsim 1/\omega$ indeed.

\clearpage
\subsection{Resonant transmission} \label{resonantsec}
\begin{figure}[!ht]
\centering
\includegraphics[width=330pt]{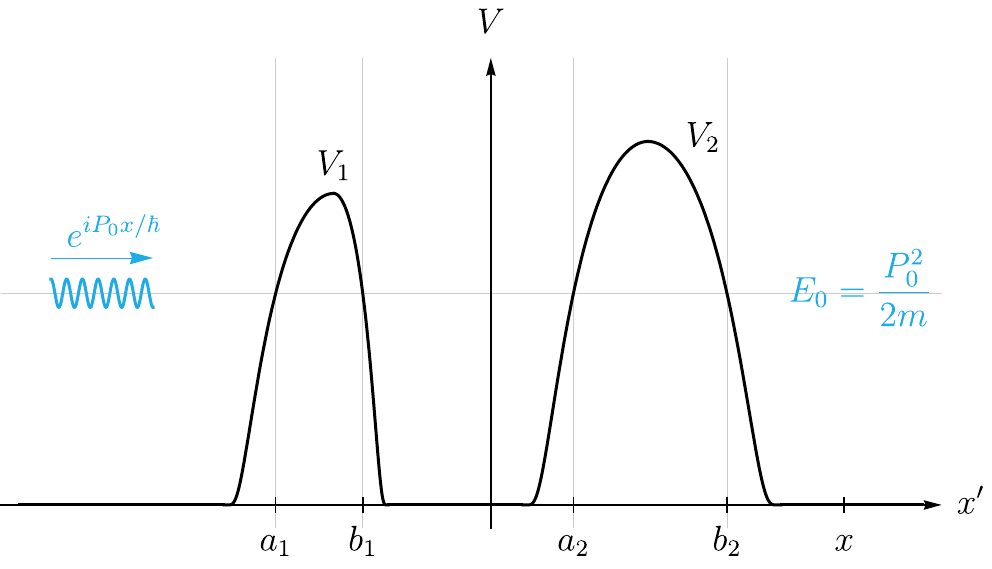}
\caption{The setup for resonant transmission: two features, with the energy of the incoming particle closely matching one of the WKB-quantized levels of the region in between the barriers.}
\label{resonanttunneling}
\end{figure}

Quantum mechanics allows particles to tunnel through barriers (\S\ref{underbarriersec}) or reflect above them (\S\ref{overbarriersec}). Astoundingly, particles can also exhibit \emph{unit} probability to tunnel through \emph{two} successive barriers. Resonant transmission occurs when the incident energy matches a resonant eigenstate of the intermediate region to exponential precision (Fig.~\ref{resonanttunneling}).
We will derive this resonant enhancement from the path integral by combining the complex-time contours of \S\ref{underbarriersec} with the multi-bounce summation of \S\ref{decaysec}. Following the formalism of \S\ref{mainsec}, we compute the transmission of a right-moving energy eigenstate incident from the left ($x_0<a_1$) with energy $E_0 = P_0^2/2m$:
\begin{equation} \label{initialstate_resonant}
    \psi(x_0,0) = \frac{\mathcal{N}_0}{\sqrt{P_0}} \exp \left( \frac{i}{\hbar} P_0(x_0-X_0)\right) \,,
\end{equation}
where $X_0$ is a reference point in the free region to the left of the double feature.

\begin{figure}[!ht]
\centering
\includegraphics[width=240pt]{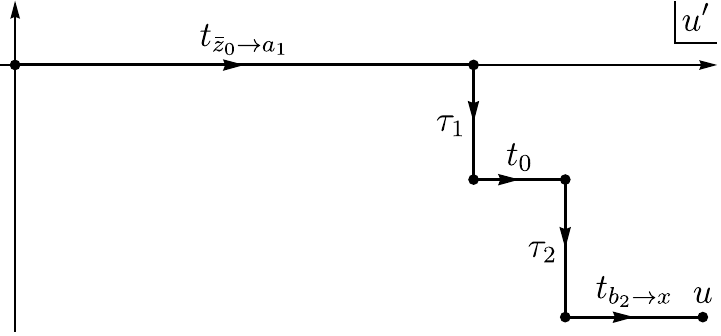}
\hfill
\includegraphics[width=240pt]{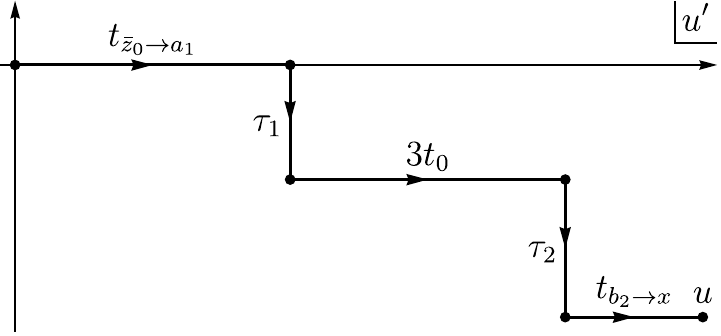}
\caption{Left: the base trajectory ($\ell_0 = \ell_1 = \ell_2 = 0$) contributing to the transmitted wave function $\psi(x,t)$ in the rightmost free region. The particle starts at $\bar{z}_0 < a_1$ with real momentum $P_0$. It travels to the first barrier in real time $t_{\bar{z}_0\to a_1}$, tunnels through it via a negative imaginary-time step $-i\tau_1$, and rolls across the classically allowed central region in real time $t_0$. After a second tunneling step $-i\tau_2$, it escapes to $x$ in real time $t_{b_2 \to x}$, reaching the final complex time $u$. We analytically continue by letting $u \to t$. This is the unique path of its kind.
Right: a resonant trajectory featuring additional real-time bounces in the central region (shown here for $\ell_0 = 1$).}
\label{nL=0FIG}
\end{figure}

\begin{figure}[!ht]
\centering
\includegraphics[width=300pt]{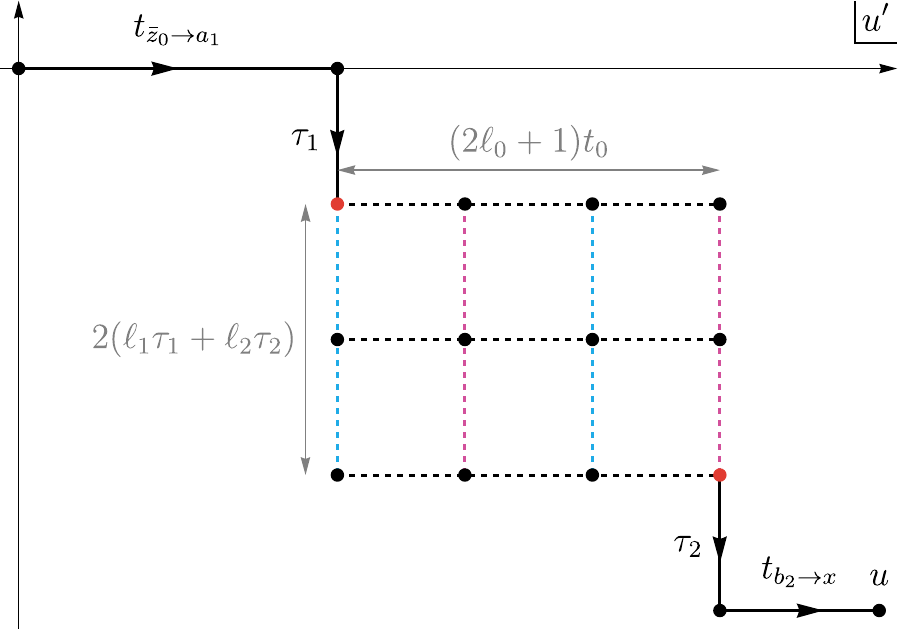}
\caption{The general solution. The part of the solution that is common to all is shown in black. But now, each time the left WKB turning point $b_1$ is reached, one can go through the barrier and back, a total of $\ell_1$ times. These $\ell_1$ bounces can only occur at the left turning point (i.e.\ along the blue dashed lines); they are each weighted by $e^{-2S_1/\hbar}$. Likewise for the other $\ell_2$ bounces which start at $a_2$; they may only occur along the pink dashed lines and are weighted by $e^{-2S_2/\hbar}$. The number of such paths is given in \eqref{amountofpaths}.}
\label{nL=l0_nE1=l1_nE2=l2FIG}
\end{figure}

\paragraph{Path integral computation}
We will compute the wave function $\psi(x,u)$ to the right of the barrier by summing over real-valued classical trajectories on the taxicab time contours shown in Fig.~\ref{nL=l0_nE1=l1_nE2=l2FIG}, which can be constructed starting from the \emph{base solution} to the equations of motion that starts at $\bar{z}_0$ and reaches $x$ at time $u$ by penetrating straight through the potential barriers, as shown in Fig.~\ref{nL=0FIG}  (which we will denote by $\ell_0=\ell_1=\ell_2=0$). The base contribution is computed identically to the transmitted wave function in \S\ref{underbarriersec}, after continuing $u\to t$:
\begin{equation}
    \psi_{\textsf{base}}(x,t)= \frac{\mathcal{N}_0}{\sqrt{k_0(x)}}\exp \left[ \frac{i}{\hbar} \left( \int_{X_0}^{a_1} + \int_{b_2}^{x} \right) k_0  + i \varphi \right] e^{-(S_1+S_2)/\hbar} e^{-i E_0 t/\hbar}\,,
\end{equation}
where we introduced the barrier integrals
\begin{equation}
    \varphi=\frac{1}{\hbar}\int_{b_1}^{a_2}k_0\,,\quad S_1 \equiv \int_{a_1}^{b_1} \kappa_0 \,, \quad S_2 \equiv \int_{a_2}^{b_2} \kappa_0 \,.
\end{equation}
All other contributing trajectories are labeled by bounce multiplicities $(\ell_0, \ell_1, \ell_2)$: $\ell_0$ denotes the number of additional real-time bounces back and forth in the middle region, while $\ell_1$ and $\ell_2$ are the additional bounces through the first and second barriers, respectively (see Fig.~\ref{nL=l0_nE1=l1_nE2=l2FIG}).
These solutions start at initial points given by
\begin{equation}\label{eq:initiall0l1l2}
    \bar{z}_0(\ell_0,\ell_1,\ell_2)=\bar{z}_0(0,0,0)+2 \ell_0\frac{P_0 t_0}{m} \,, \quad t_0 \equiv t_{b_1 \to a_2} \,.
\end{equation}
Analogously to \S\ref{decaysec}, we find that each of these solutions gives a contribution that can be related to the base wave function after the analytic continuation $u^{(\ell_0,\ell_1,\ell_2)} \to t$, as
\begin{equation}
    \psi_{\ell_0,\ell_1,\ell_2}(x,t)= \psi_{\textsf{base}}(x,t) \left(- e^{2 i \varphi} \right)^{\ell_0} \left(-\frac{1}{2}e^{-2 S_1/\hbar} \right)^{\ell_1} \left(-\frac{1}{2} e^{-2 S_2/\hbar} \right)^{\ell_2} \,.
\end{equation}
The expression above comprises both the semiclassical exponent and the one-loop factor, where we already accounted for the factors of $1/2$ associated with the bounces. Moreover, analogous combinatorics to that of \S\ref{insidesec} shows that the total number of distinct solutions characterized by $(\ell_0, \ell_1, \ell_2)$ is given by
\begin{equation} \label{amountofpaths}
    C(\ell_0, \ell_1, \ell_2) = \binom{\ell_0+\ell_1}{\ell_1} \binom{\ell_0+\ell_2}{\ell_2} \,.
\end{equation}

Defining $\Upsilon_i \equiv e^{-2 S_i/\hbar}$, we obtain the total wave function\footnote{To be precise, one can only consistently compute the transmission by summing up to a maximum value $\ell_0$ that ensures that all the initial positions from \eqref{eq:initiall0l1l2} are located to the left of the potential. In what follows, we will always consider times sufficiently large so that the sum below has converged and we can neglect this subtlety. The requirement can be shown to be $(t - m x/P_0)/t_0 \gg e^{2S_1/\hbar}, e^{2S_2/\hbar}$.}
\begin{align}
    \psi(x,t) &= \psi_{\textsf{base}}(x,t) \sum_{\ell_0 = 0}^{\infty} \left( -e^{2 i \varphi} \right)^{\ell_0} \left[ \sum_{\ell_1=0}^\infty \binom{\ell_0+\ell_1}{\ell_1} \left( -\frac{1}{2} e^{-2 S_1/\hbar} \right)^{\ell_1} \right] \times \left[ 1 \leftrightarrow 2 \right] \times \left[ 1 + \mathcal{O}(\hbar) \right] \, \notag \\
    &= \psi_{\textsf{base}}(x,t)\sum_{\ell_0 =0}^{\infty} \left[ \frac{-e^{2 i \varphi}}{\left(1 + \Upsilon_1/2 \right)\left(1 + \Upsilon_2/2 \right)} \right]^{\ell_0} \left[ 1 + \mathcal{O}(\hbar) \right] \notag \\
    &= \frac{\psi_{\textsf{base}}(x,t)}{1 + e^{2 i \varphi} \left[ 1 - (\Upsilon_1+\Upsilon_2)/2 \right]} \left[ 1 + \mathcal{O}(\hbar) \right] \,,
\end{align}
where we neglected exponentially suppressed corrections at every step.

We see that the transmitted pulse is highly suppressed by the baseline amplitude $\psi_{\textsf{base}} \sim \sqrt{\Upsilon_1 \Upsilon_2}$ \emph{unless} the denominator is also exponentially small. This resonant enhancement is maximized when the phase satisfies the WKB quantization condition for the central well
\begin{equation}
    e^{2 i \varphi} = -1 \quad\Longleftrightarrow\quad \int_{b_1}^{a_2} k_0 = \pi \hbar \left( n + 1/2 \right)\,.
\end{equation}
Small deviations from resonance can be parametrized by $\delta \equiv \varphi(E_0) - \pi \left(n + 1/2 \right)\ll 1$. We find
\begin{equation}
    \psi(x,t) = \frac{2\sqrt{\Upsilon_1\Upsilon_2}}{\Upsilon_1 + \Upsilon_2 - 4 i \delta} \frac{\mathcal{N}_0}{\sqrt{k_0(x)}}\exp \left[ \frac{i}{\hbar} \left( \int_{X_0}^{a_1} + \int_{b_2}^{x} \right) k_0  + i \pi \left( n + 1/2 \right) \right] e^{-i E_0 t/\hbar} \left[ 1 + \mathcal{O}(\hbar) \right] \,.
\end{equation}
We can now isolate the effective resonant transmission amplitude $\mathcal{T}$:
\begin{equation} \label{resonant_amplitude}
    \mathcal{T} \approx \frac{2 \sqrt{\Upsilon_1 \Upsilon_2}}{\Upsilon_1 + \Upsilon_2 - 4 i \delta} = \frac{2 \sqrt{\Upsilon_1 \Upsilon_2}}{\Upsilon_1 + \Upsilon_2} \left[ 1 + \left( \frac{4 \delta}{\Upsilon_1 + \Upsilon_2} \right)^2 \right]^{-1/2} e^{i \theta_\textsf{res}} \,,
\end{equation}
where the transmitted wave acquires a resonant phase shift $\theta_\textsf{res}$ given by
\begin{equation}
    \theta_\textsf{res} = \tan^{-1} \left( \frac{4 \delta}{\Upsilon_1 + \Upsilon_2} \right) \,.
\end{equation}
Squaring the amplitude $\mathcal{T}$ yields the familiar Breit--Wigner profile for the transmission probability,
\begin{equation} \label{resonant_probability}
    \abs{\mathcal{T}}^2 \approx \frac{4 \Upsilon_1 \Upsilon_2}{(\Upsilon_1 + \Upsilon_2)^2 + 16 \delta^2} \,.
\end{equation}
This demonstrates the resonant enhancement: exactly at resonance ($\delta = 0$), the transmission probability is $\abs{\mathcal{T}}^2 = 4 \Upsilon_1 \Upsilon_2/(\Upsilon_1 + \Upsilon_2)^2$. If the barriers are strictly symmetric such that their under-barrier actions are equal, the numerator precisely cancels the denominator. The particle tunnels through both barriers with unit probability, a consequence of the fine-tuned constructive interference of the bounces. For unequal barriers, the transmission coefficient is enhanced from $e^{-(S_1 + S_2)/\hbar}$ to $e^{-\abs{S_1 - S_2}/\hbar}$.

\paragraph{About wave packets}
Observing resonant transmission in a wave packet requires the energy spread to fit entirely within the exponentially narrow resonance band of width $\Gamma = \hbar/t_0 \times (\Upsilon_1 + \Upsilon_2)$. For a Gaussian initial state such as \eqref{eq:gaussian_wavepacket1}, this dictates an extreme spatial constraint $\sigma_x = \sqrt{\hbar/\alpha} \gg (P_0 t_0/ m) \exp \left( 2 \min \{ S_1, S_2 \} \right)$, so the incident wave packet must be exponentially broader than the double-barrier system itself. Furthermore, the transmitted wave packet experiences a substantial time delay $t_\textsf{offset} = \hbar/\Gamma$, corresponding to the lifetime of the bound state excited inside the well. This is why we, unfortunately, will always have to go through the cold shower section before reaching the swimming pool.

\section{Conclusion}
We considered the time evolution of semiclassical states in quantum mechanics from the path integral approach. Approximating the path integral by contributions from the vicinity of saddle points, we arrived at \eqref{masterformula} which accounts for the next-to-leading order behavior as $\hbar \to 0$. An overall spacetime-independent constant $\mathcal{N}_P$ accompanying subleading saddle points at early times was left undetermined. A sharply defined question of when saddle points indeed contribute to the time-evolved state was set up, but it was not solved in the general case. In this \enquote{direct} calculation, time is kept on the real axis throughout but the saddle points are, generally, complex-valued paths.

In \S\ref{examplessec} we illustrated the formalism in a collection of classic problems in one-dimensional quantum mechanics. Coherent states in the harmonic oscillator (\S\ref{HOsec}) provide a simple setup in which complex-valued paths in real time arise. For the transmission of particles with insufficient energy to classically make it across a barrier (\S\ref{underbarriersec}), instead, we solved the equations of motion in the complex time plane. The endpoint of time was kept as a free complex variable, which was analytically continued to the real axis at the end of the computation. In this \enquote{indirect} approach, the saddle point solutions remain real-valued along taxicab contours in the complex time plane. This approach was also used in the description of the decay of a general, excited metastable state (\S\ref{decaysec}). The famous \enquote{bounce} \cite{Coleman:1977py} \textdash a zero-energy solution in infinite imaginary time \textdash finds its finite-energy and finite-time counterpart here in solutions which oscillate in real time in the metastable region and tunnel out in finite time by imaginary time evolution in the inverted barrier. The exponential decay of probability inside the metastable region arises from a combinatorial factor that counts the multiplicity of solutions with a given number of bounces. In \S\ref{numericssec} we compared the direct and indirect methods with the numerical solution to the Schr\"odinger equation in an explicit tunneling setup. We identified interesting exchanges of dominance between complex saddle points and confirmed them with the numerics.

In \S\ref{overbarriersec} and \S\ref{resonantsec} we showed how over-barrier reflection and resonant transmission are treated in this formalism. In all examples, assuming every relevant saddle point was found, the path integral calculations were shown to match known WKB results.

\vspace*{\baselineskip}
\noindent We close with a speculation and an open question.

The speculation is that $\mathcal{N}_P = 1$ for all contributing saddle points in \eqref{masterformula}, except when this is not the case. More precisely, in the direct method of \S\ref{mainsec} in which time is real and solutions are complex-valued, half of the authors believe $\mathcal{N}_P = 1$ always. Their argument is that the path integral measure \eqref{PImeasure} should be common to all saddle points, and we have determined $\mathcal{N}_P = 1$ for the special saddle that is continuously connected to the identity as $t \to 0$. Furthermore, $\mathcal{N}_P$ is consistent with one for the handful of non-trivial saddle points in the \textdash it would appear \textdash arbitrary numerical example of \S\ref{numericssec}. The other half of the authors argue that $\mathcal{N}_P = 1/2^\ell$ for solutions with $\ell$ bounces in order for them to correctly describe the decay of a metastable state at late times, so $\mathcal{N}_P$ cannot always be one.\footnote{One fourth of the authors are also puzzled about why one should sum over different complex time representations of the \emph{same} solution $\bar{z}(u')$, which gives the combinatorial factor \eqref{combinatorialfactor}. The other three-fourths suggest that one should sum over one representative per homotopy class $\{ u' = \gamma(l) \colon 0 \to u \}$ in $\mathbb{C} \backslash \{ \text{turning point times} \}$ (cf.~\cite{Witten:2021nzp,Turiaci:2025xwi}). We leave the problem of how the continuation back to real time should be done, based on the singularity structure of $\bar{z}(u')$, to future work.} The first half insists that this must be an artifact of the indirect method that analytically continues time, which we have not fully understood. We warmly welcome the intervention of a Galileo among our readers to resolve this dialogue.

The open question is whether there exists a powerful yet practical,\footnote{I.e.\ less trivial than the one following from \eqref{simpler0condition} but that does not require solving the problem around \eqref{downwardflowEq}, which is generally difficult.} sufficient criterion to \emph{rule out} the contribution of complex saddle points to \eqref{mainPI}. In quantum gravity, the equivalent of such a criterion was proposed in \cite{Witten:2021nzp}. Clearly, there is something the present authors still do not understand about one-dimensional quantum mechanics, let alone about gravitation.

\section*{Acknowledgements}
We thank Matt Kleban, Simon Krekels, Cammie Norton, Jo\~ao Penedones, Riccardo Rattazzi, S\'ebastien Reymond, Sergey Sibiryakov and Nils Wagner for useful discussions.
J.K.~is supported by the Research Foundation - Flanders (FWO) doctoral fellowship 1171825N, the KU Leuven grant C16/25/010, the STFC consolidated grant ST/T000694/1 and the Simons Investigator award \#620869.
F.R.~is supported by the research grant number 20227S3M3B \enquote{Bubble Dynamics in Cosmological Phase Transitions} under the program PRIN 2022 of the Italian Ministero dell'Università e della Ricerca (MUR).

\phantomsection
\addcontentsline{toc}{section}{\refname}
\bibliographystyle{klebphys2}
\bibliography{refs}
\end{document}